\documentclass[twocolumn,prb,showpacs,preprintnumbers,amsmath,eqsecnum,amssymb]{revtex4}\usepackage{graphicx}
\usepackage{dcolumn}% Align table columns on decimal point
\usepackage{bm}% bold math

\begin{document}

\title{Spin diffusion
and injection in semiconductor
structures: Electric field effects}

\author{Z. G. Yu and M. E. Flatt\'e}

\address{Department of Physics and Astronomy, University of Iowa,
Iowa City, Iowa 52242}
\date{\today}

\begin{abstract}
In semiconductor spintronic devices, the semiconductor
is usually lightly doped and nondegenerate, and moderate
electric fields can dominate the carrier motion.
We recently
derived a drift-diffusion equation for spin polarization in the
semiconductors by consistently taking into account electric-field
effects and nondegenerate electron statistics and identified a
high-field diffusive regime which has no analogue in metals.
Here
spin injection from a ferromagnet (FM) into a nonmagnetic
semiconductor (NS)
is extensively studied by applying this  spin drift-diffusion
equation to several typical injection structures such as
FM/NS, FM/NS/FM, and FM/NS/NS structures.
We find that in the high-field regime
spin injection from a ferromagnet  into a
semiconductor is enhanced by several orders of magnitude.
For injection structures with
interfacial barriers, the electric field further enhances spin injection
considerably. In FM/NS/FM structures high electric fields destroy
the symmetry between the two magnets at low fields, where
both magnets are equally important for spin injection, and spin
injection becomes locally determined by the
magnet from which carriers flow into the semiconductor.
The field-induced spin injection enhancement should also be
insensitive to the presence of a highly doped nonmagnetic semiconductor
(NS$^+$)
at the FM interface, thus FM/NS$^+$/NS structures should also manifest
efficient spin
injection at high fields. Furthermore, high fields substantially reduce the
magnetoresistance observable in a recent experiment on spin injection
from magnetic semiconductors.
\end{abstract}
 \phantom{.}

\maketitle

\section{introduction}

Semiconductor devices based on the control and manipulation
of electron spin (semiconductor spintronics)
have recently attracted considerable attention
since the discovery of
long spin relaxation times and large spin
transport distances in semiconductors and various device
structures.\cite{wolf,spin_book}
In order to design and fabricate high-performance
spintronic devices, a comprehensive understanding of
spin transport and injection properties of semiconductors
and heterostructures is needed.

In theoretical
studies of spin transport and injection in semiconductors \cite{schmidt,
rashba, smith,fert}
the spin polarization is usually assumed to obey the same diffusion
equation as in metals,\cite{son}
\begin{equation}
\nabla^2 (\mu_{\uparrow}-\mu_{\downarrow})
-(\mu_{\uparrow}-\mu_{\downarrow})/L^2=0,
\label{deg}
\end{equation}
where $\mu_{\uparrow(\downarrow)}$ is the electrochemical potential
of up-spin (down-spin) electrons.
In this diffusion equation, the
electric field does not play any role, and spin polarization
decays away on a length scale of $L$ from an injection point.
This is reasonable
for metals because
the electric field {\bf E} is essentially screened.
For semiconductor spintronic devices, however,
the semiconductor often
is lightly doped and nondegenerate, and a moderate electric field
can dominate the carrier motion. In fact, experiments
have shown that
electric fields can affect spin diffusion in semiconductors
dramatically.\cite{length,malajovich}

In Ref. \onlinecite{us},
we examined the role of electric field on spin transport
in nondegenerate
semiconductors and derived a drift-diffusion equation
for spin polarization,
\begin{equation}
\nabla^2 (n_{\uparrow}-n_{\downarrow})+\frac{e{\bf E}}{k_B T} \cdot \nabla
(n_{\uparrow}-n_{\downarrow})
-\frac{n_{\uparrow}-n_{\downarrow}}{L^2}=0,
\label{nondeg}
\end{equation}
where $n_{\uparrow(\downarrow)}$ is the deviation of
up-spin (down-spin) electron density from its equilibrium value,
$k_B$ the Boltzmann constant, and $T$ the temperature.
This equation consistently takes into account
electric-field effects and nondegenerate electron statistics.
We identified a high-field diffusive regime which has no analogue
in metals. This regime occurs for field as small as 1 V/cm at low
temperatures.
Two distinct
spin diffusion lengths now characterize
spin motion, i.e., up-stream ($L_u$) and down-stream ($L_d$)
spin diffusion lengths.
This is a further example of the analogy \cite{packet,unipolar} between
up/down-spin electron
in semiconductor spin transport
and majority/minority carriers  in semiconductor charge transport, where
the presence of electric field also results in two charge diffusion
lengths.\cite{book}
We applied this  spin drift-diffusion equation
to study spin injection
from a ferromagnet into a semiconductor, and
showed that the electric-field effects on spin injection could be
described in terms of the two field-induced spin diffusion lengths
in the semiconductor.

In this paper, first, we derive a more general drift-diffusion equation of
spin polarization valid for both doped semiconductors and metals,
and
demonstrate the development from Eq. (\ref{nondeg}) to Eq. (\ref{deg}) as
the
system changes from nondegenerate to degenerate.
We establish the relation between the electrochemical potential
splitting and density imbalance of up-spin and down-spin electrons
in nondegenerate systems, and clarify the two spin polarizations, i.e.,
spin polarization of current and
spin polarization of density.
We then use  the  spin drift-diffusion equation for nondegenerate
systems, Eq. (\ref{nondeg}),
to analyze several typical one-dimensional device geometries.
We find that high fields also
enhance spin injection from a ferromagnet
to a semiconductor
in structure with a spin-selective
barrier.
Rashba,\cite{rashba} Smith and Silver,\cite{smith}
Fert and Jaffr\`es,\cite{fert}
and Flatt\'e, Byers, and Lau in Ref. \onlinecite{spin_book}
have considered such a barrier in the low-field
regime. The field enhancement and the interface enhancement of
spin injection may reinforce each other to achieve high injection
efficiencies in different structures.

Next, we study spin injection in sandwiched
FM/NS/FM structures with and without spin-selective interfacial
barriers. At low fields the two magnets are equally important
to determine spin injection into the semiconductor and the spin injection
efficiency is sensitive to the relative orientation of the two magnets. We
find that in the high-field regime, this symmetry is broken and spin
injection
is locally determined by the magnet from which carriers are
injected into the semiconductor. The spin injection efficiency
can be enhanced by orders of magnitude by increasing the electric field
for both parallel and anti-parallel orientation of the two magnets.

We further consider FM/NS$^+$/NS structures, where a highly doped
nonmagnetic semiconductor (NS$^+$)
is
placed near the magnet interface. Such a configuration is common
in structures designed to overcome the Schottky barrier between a magnet and
a semiconductor, and is intrinsic
to FM/InAs, where  densely occupied surface states form at the interface.
We find that spin injection at the strong-field limit in such a structure
is controlled by the total electric
current flowing into semiconductors and insensitive to
the  distinction between semiconductors. Thus high fields can effectively
enhance spin injection in such structures as well.

Finally we explore electric-field effects on magnetoresistance of a magnetic
semiconductor (MS)/NS/MS structure. A
magnetic semiconductor
can have extremely large spin polarization at low
temperatures and the magnetization can be easily adjusted by applying an
external magnetic field.
A  large positive magnetoresistance
have been observed in MS/NS/MS structures.\cite{magnetoresistance}
We find that this magnetoresistance
collapses in the high-field regime, suggesting a sensitive test of the
electric-field effects on spin transport in semiconductors.

The paper is organized as follows. In Sec. II we review the  general
spin drift-diffusion equation in nondegenerate and degenerate
systems and analyze the field-induced
spin diffusion lengths. In Sec. III we investigate spin injection
in FM/NS structures with an interfacial barrier. Sections IV and V contain
results on spin injection in FM/NS/FM structures and in FM/NS/NS structures,
respectively.
Section VI is devoted to the electric-field effects on magnetoresistance in
MS/NS/MS
structures. In Sec. VII, we summarize our conclusions.

\section{Electric field and spin transport}

In this section we derive a more general drift-diffusion equation for spin polarization
valid in both degenerate and nondegenerate systems and
discuss the electric-field effects on spin transport by analyzing the structure of
this equation. Discussion of the drift-diffusion equations of carrier motion in semiconductors
involving spin-dependent processes can be tracked back decades ago. Pierce {\it et al.}
incorporated spin relaxation in a one-dimension diffusion model of up-spin and down-spin
carrier densities at zero electric field.\cite{pierce} Sogawa {\it et al.}
investigated spin transport in wires with a set of
complete drift-diffusion equations for minority carriers, which explicitly include spin-flip and
recombination processes as well as the electric field.\cite{sogawa}
More recently \^Zuti\'c {\it et al.} displayed such
equations for both minority and majority carriers.\cite{zutic}
 Usually extensive
 numerical calculations are required to solve the set of
drift-diffusion equations together with the Poisson's equation self-consistently.
The spin drift-diffusion equation we  derive is  a {\em single}
equation instead of a set of equations and can be solved analytically
in several interesting geometries.
The role of this spin drift-diffusion equation in spin transport
is similar to that of the ambipolar drift-diffusion equation
in charge transport.

\subsection{Drift-diffusion equation for spin polarization}

The system we consider here is
$n$-doped ($p$-doped systems can be analyzed similarly),
which can be ferromagnetic or nonmagnetic. The analysis
presented in this section is valid not only in doped semiconductors
but also in metals.
We assume that there
is no space charge and the material
is homogeneous.
The current for up-spin and down-spin can be written as
\begin{subequations}
\begin{eqnarray}
{\bf j}_{\uparrow}&=&\sigma_{\uparrow}{\bf
E} +e D_{\uparrow} {\bf \nabla} n_{\uparrow},\\
{\bf j}_{\downarrow}&=&\sigma_{\downarrow}{\bf
E} +e D_{\downarrow} {\bf \nabla} n_{\downarrow},
\end{eqnarray}
\end{subequations}
which consists of the drift current and the diffusion one.
Here $D_{\uparrow(\downarrow)}$ is the up-spin (down-spin) electron
diffusion constant
and $\sigma_{\uparrow(\downarrow)}$ the up-spin (down-spin)
conductivity.
The change of up-spin (down-spin) conductivity from its unperturbed
value $\sigma^0_{\uparrow(\downarrow)}$ in the presence of spin
polarization,
$\Delta \sigma_{\uparrow(\downarrow)} \equiv \sigma_{\uparrow(\downarrow)}
-\sigma^0_{\uparrow(\downarrow)}$, is assumed to be
proportional to $n_{\uparrow(\downarrow)}$,
the up-spin (down-spin) electron density {\em deviation}
from its equilibrium
value $n^0_{\uparrow(\downarrow)}$,
\begin{equation}
\Delta
\sigma_{\uparrow(\downarrow)}=n_{\uparrow(\downarrow)}e\nu_{\uparrow(\downarrow)}.
\end{equation}
Here the mobility $\nu_{\uparrow(\downarrow)}$ is independent of
field and density over the range of density variation
$n_{\uparrow(\downarrow)}$.

The continuity equations for up-spin and down-spin electrons in systems
including
spin-flip scattering process
are
\begin{eqnarray}
\frac{\partial n_{\uparrow}}{\partial t}&=&
-\frac{n_{\uparrow}}{\tau_{\uparrow\downarrow}}+\frac{n_{\downarrow}}
{\tau_{\downarrow\uparrow}}+\frac{1}{e}{\bf \nabla}\cdot{\bf
j}_{\uparrow},\\
\frac{\partial n_{\downarrow}}{\partial t}&=&
-\frac{n_{\downarrow}}{\tau_{\downarrow\uparrow}}+\frac{n_{\uparrow}}
{\tau_{\uparrow\downarrow}}+\frac{1}{e}{\bf \nabla}\cdot{\bf
j}_{\downarrow},
\end{eqnarray}
where $\tau^{-1}_{\uparrow\downarrow}$ ($\tau^{-1}_{\downarrow\uparrow}$) is
the
rate with which up-spin (down-spin) electrons scatter to down-spin (up-spin)
electrons.
Here the recombination process is neglected because the system we consider
is doped
(unipolar).
In steady state [$\partial n_{\uparrow(\downarrow)}/\partial t=0$], we have
\begin{equation}
{\bf
\nabla}\sigma_{\uparrow}\cdot{\bf E} +\sigma_{\uparrow}{\bf
\nabla}\cdot{\bf E}+e D_{\uparrow}\nabla^2n_{\uparrow}
=\Big(\frac{n_{\uparrow}}{\tau_{\uparrow\downarrow}}-\frac{n_{\downarrow}}{\tau_{\downarrow\uparrow}}
\Big)e,
\label{upspin}
\end{equation}
\begin{equation}
{\bf
\nabla}\sigma_{\downarrow}\cdot{\bf E} +\sigma_{\downarrow}{\bf
\nabla}\cdot{\bf E}+e D_{\downarrow}\nabla^2n_{\downarrow}
=\Big(\frac{n_{\downarrow}}{\tau_{\downarrow\uparrow}}
-\frac{n_{\uparrow}}{\tau_{\uparrow\downarrow}}\Big)e,
\label{downspin}
\end{equation}
where ${\bf \nabla}\cdot{\bf E} = -e(n_{\uparrow}+n_{\downarrow})
/\epsilon$, and
$\epsilon$ is the dielectric constant of the
system.

For a homogeneous
system without space-charge,
$n_{\uparrow}+n_{\downarrow}$ should be balanced by a local change
of hole concentration. In doped systems, however,
spin polarization can be created without changing
electrons or hole densities,\cite{neutrality,packet} and therefore,
\begin{equation}
n_{\uparrow}+ n_{\downarrow}=0.
\label{neutral}
\end{equation}

Care is required, however, to avoid setting ${\bf \nabla}\cdot{\bf E}=0$
directly
in Eqs. (\ref{upspin}) and (\ref{downspin}).\cite{book}  Instead we multiply
Eq. (\ref{upspin}) by $\sigma_{\downarrow}$
and Eq. (\ref{downspin}) by $\sigma_{\uparrow}$, and substract one from the
other, eliminating
the terms containing ${\bf \nabla}\cdot{\bf E}$. Only then do we set
$ n_{\uparrow}+  n_{\downarrow}=0$. Now we have
\begin{equation}
\nabla^2(n_{\uparrow}-n_{\downarrow})+\frac{\nu}{eD}e{\bf E}
\cdot{\bf\nabla}(n_{\uparrow}-n_{\downarrow})
-\frac{n_{\uparrow}-n_{\downarrow}}{L^2}=0,
\label{general}
\end{equation}
where the effective mobility $\nu$ and the effective diffusion constant $D$
for spin polarization are
\begin{subequations}
\begin{eqnarray}
\nu
&=&\frac{\sigma_{\uparrow}\nu_{\downarrow}+\sigma_{\downarrow}\nu_{\uparrow}}
{\sigma_{\uparrow}+\sigma_{\downarrow}},\label{mobility}\\
%\end{equation}
%\begin{equation}
D &=&\frac{\sigma_{\uparrow}D_{\downarrow}+\sigma_{\downarrow}D_{\uparrow}}
{\sigma_{\uparrow}+\sigma_{\downarrow}}.
\label{diffusion}
\end{eqnarray}
\end{subequations}
And
\begin{equation}
L=\sqrt{D\tau_S}
\end{equation}
is the
intrinsic spin-diffusion length, where the spin relaxation time $\tau_S$ is
defined via $\tau^{-1}_S=\tau^{-1}_{\uparrow\downarrow}+
\tau^{-1}_{\downarrow\uparrow}$.
Equations (\ref{mobility}) and (\ref{diffusion}) indicate
that  the behavior of spin transport is controlled by the {\it minority}
spin species. This is analogous to the ambipolar charge transport, where
minority charge carriers
determine the behavior of excess charge transport.\cite{book}

For nonmagnetic systems, $\nu_{\uparrow}=\nu_{\downarrow}=\nu$ and
$D_{\uparrow}=D_{\downarrow}=D$. For ferromagnetic systems, $\nu$
and $D$ are approximately the mobility and the diffusion constant
for the lower-conductivity  spin species, usually the minority
spins. We will assign the down-spin label to this species, so $\nu
\simeq \nu_{\downarrow}$ and $D \simeq D_{\downarrow}$. Thus the
coefficient of the second term in Eq. (\ref{general}) can be
approximated from the single-band form of the Einstein
relation,\cite{book}
\begin{equation}
\frac{\nu}{eD}=-\int_0^{\infty}N({\cal E})\frac{\partial f_0}
{\partial {\cal E}}d{\cal E}\bigg/ \int_0^{\infty}N({\cal E})f_0({\cal E})
d{\cal E},
\label{einstein}
\end{equation}
where ${\cal E}$ is the energy measured from the bottom edge of
the conduction band for the minority spin, $N({\cal E})$ is the
density of states for the minority spin and $f_0$ the distribution
function. A more accurate evaluation of $\nu/eD$ when
$\nu_{\uparrow} \ne \nu_{\downarrow}$ could not be done without
knowledge of, e.g., $\nu_{\uparrow}/\nu_{\downarrow}$.

One special exception exists, however, for nondegenerate
semiconductors, where $f_0$ has the Boltzmann form, $f_0 \sim
e^{-{\cal E}/k_B T}$, and $\nu/eD=1/k_BT$. Thus we obtain Eq.
(\ref{nondeg}) to describe the transport of
$n_{\uparrow}-n_{\downarrow}$,
the natural measure of the spin polarization in semiconductors. We
emphasize that Eq. (\ref{nondeg}) is also valid for highly
spin-polarized (including ferromagnetic) nondegenerate
semiconductors.

For degenerate systems, $f_0$ in Eq. (\ref{einstein}) should have
the Fermi-Dirac form. In a three-dimensional (3D) system, $N({\cal
E})=A {\cal E}^{1/2}$, we have approximately
\begin{equation}
\frac{\nu}{eD}=\frac{1}{2k_B T}
\frac{F_{-1/2}[(\varepsilon_F-\varepsilon^b_{\downarrow})/k_B T]}{F_{1/2}[
(\varepsilon_F-\varepsilon^b_{\downarrow})/k_B T]},
\label{3d}
\end{equation}
where $\varepsilon_F$ is the Fermi energy,
$\varepsilon^b_{\downarrow}$ the bottom edge of the conduction
band for the minority spin, and $F_n(\xi)=\int_0^{\infty}dx~ x^n
[e^{x-\xi}+1]^{-1}$.\cite{book} In a two-dimensional (2D) system,
$N({\cal E})$ is a constant,
\begin{equation}
\frac{\nu}{eD}=\frac{1}{k_B T
F_0[(\varepsilon_F-\varepsilon^b_{\downarrow})/k_BT][1+e^{-(\varepsilon_F-\varepsilon^b_{\downarrow})/k_B
T}]}. \label{2d}
\end{equation}
We can define a critical field
\begin{equation}
E_c \equiv \frac{1}{eL}\Big(\frac{\nu}{eD}\Big)^{-1},
\end{equation}
such that when $E>E_c$, the drift term will be more important than
the diffusive term in Eq. (\ref{general}), and neglecting the
electric-field effects on spin transport in this regime cannot be
justified. In Fig. 1, we plot $E_c$ as a function of electron
density for different temperatures in 2D and 3D $n$-doped GaAs
using a typical spin diffusion length $L=2~\mu$m.\cite{length} We
can see that for electron densities ranging from $10^{15}$  to
$10^{18}$ cm$^{-3}$ the critical field $E_c$ is not beyond
realistic fields under which spintronic devices operate. In
particular, at low temperatures $E_c$ can be as low as 1 V/cm in
lightly and moderately doped semiconductors. Even  for 100\% spin
polarized {\it n}-doped ZnMnSe, $n_0=10^{18}$ cm$^{-3}$, the drift
term is relevant for $E > 200$ V/cm at $T < 30$ K. Thus the
electric field should be taken into account to  properly interpret
phenomena involving spin transport in both magnetic and
nonmagnetic semiconductors.
\begin{figure}
\vspace{10pt}\includegraphics[width=7cm]{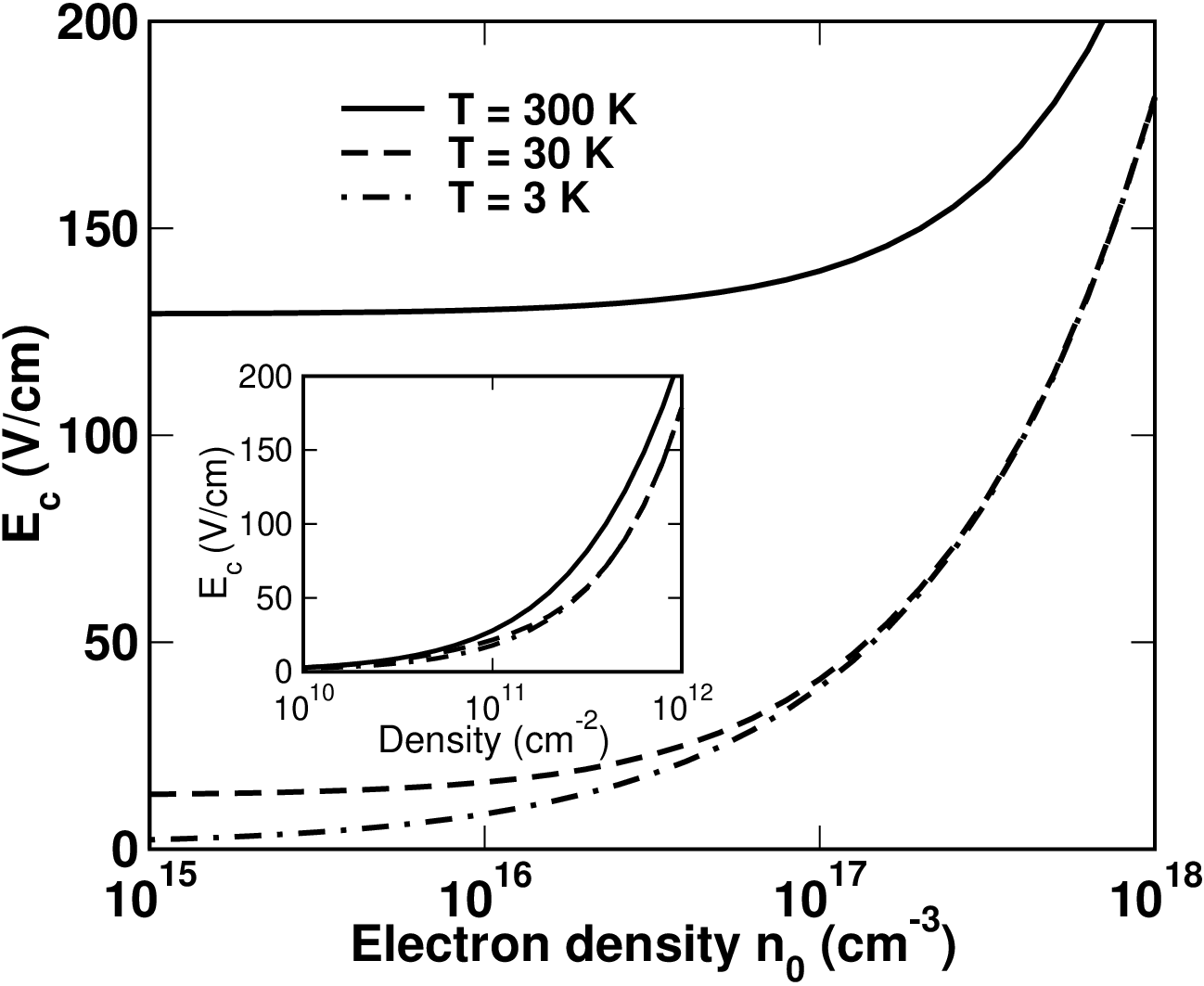}
\caption{Critical field $E_c$ as a function of electron
density
for different temperatures in 3D systems.
The inset is for 2D systems.
The effective electron
mass $m^*=0.067~m_0$,
where $m_0$ is the free
electron mass.}
\end{figure}

For highly degenerate systems such as metals, in which
$\varepsilon_F-\varepsilon^b_{\downarrow} \sim 10$ eV $\gg k_B T$,
from Eqs. (\ref{3d}) and (\ref{2d}), $\nu/eD=3/2
(\varepsilon_F-\varepsilon^b_{\downarrow})$ for a 3D system and
$\nu/eD=1/(\varepsilon_F-\varepsilon^b_{\downarrow})$ for a 2D system.  For
a typical spin diffusion length of metal,
$L \sim 100$ nm, the field has to exceed $10^6$ V/cm for
the drift term to become comparable to the diffusion term in Eq.
(\ref{general}). Thus in metals
under realistic fields
the drift term can be neglected. Using the relation between the
electrochemical
potential, $\mu_{\uparrow(\downarrow)}$, and the
nonequilibrium  carrier density, $n_{\uparrow(\downarrow)}$,
in a highly degenerate system
\begin{equation}
n_{\uparrow(\downarrow)}=eN_{\uparrow(\downarrow)}(\varepsilon_F)[\mu_{\uparrow(\downarrow)}+e\psi],
\end{equation}
where $N_{\uparrow(\downarrow)}(\varepsilon_F)$ is the up-spin (down-spin)
density of states  at the Fermi energy, and {\bf E}$=-\nabla \psi$,  we find
that
Eqs. (\ref{neutral}) and (\ref{general}) reduce to
\begin{equation}
\nabla^2\left ( \begin{array}{c}
\mu_{\uparrow} \\
\mu_{\downarrow}
\end{array} \right )
=\frac{1}{L^2}\left ( \begin{array}{cc}
{\displaystyle
\frac{\sigma_{\downarrow}}{\sigma_{\uparrow}+\sigma_{\downarrow}}} &
{\displaystyle
-\frac{\sigma_{\downarrow}}{\sigma_{\uparrow}+\sigma_{\downarrow}} } \\
{\displaystyle
-\frac{\sigma_{\uparrow}}{\sigma_{\uparrow}+\sigma_{\downarrow}} }&
{\displaystyle
\frac{\sigma_{\uparrow}}{\sigma_{\uparrow}+\sigma_{\downarrow}} }
\end{array} \right )
\left ( \begin{array}{c} \mu_{\uparrow} \\ \mu_{\downarrow}
\end{array} \right ),
\label{e_potential}
\end{equation}
which is consistent with Eq. (2.18) in Ref.
\onlinecite{hershfield}, the spin transport equation derived for
metallic systems. It is straightforward to see that Eq.
(\ref{deg}) is contained in Eq. (\ref{e_potential}). In deriving
Eq. (\ref{e_potential}) we have assumed that the spin-dependent
conductivity is proportional to the spin-dependent density of
states at the Fermi level,
$\sigma_{\uparrow}/\sigma_{\downarrow}=N_{\uparrow}(\varepsilon_F)/N_{\downarrow}(\varepsilon_F)$.
The effects of electron-electron interaction (alterations of the
spin stiffness, spin drag, etc.) will modify the value of
$\nu/eD$, although for the temperatures of greatest interest and
moderate density the corrections are small.\cite{vignale}

Equation (\ref{nondeg}) provides a framework to understand spin transport in
nondegenerate semiconductors, and Eqs. (\ref{general}-\ref{2d}) in
moderately degenerate semiconductors.
We see that  the electric field, unlike that
in metals, plays a central role on spin transport.
Comparing Eq. (\ref{nondeg}) with the drift-diffusion equation of minority
carriers (electrons)  in $p$-doped
semiconductors,\cite{book}
\begin{equation}
\nabla^2 (n-n_0)+\frac{e {\bf E}}{k_B T}\cdot \nabla
(n-n_0)-\frac{n-n_0}{L^2_e}=0,
\label{charge}
\end{equation}
where $L_e$ is the intrinsic electron diffusion length,
we find that Eq. (\ref{nondeg}) and
Eq. (\ref{charge}) have the same structure and the  electric field is
expected to play a similar
role in both situations.
For minority carrier transport it is well-known that the electric field
gives rise to two distinct
charge diffusion lengths
and considerably modifies minority charge injection.

\subsection{Spin diffusion lengths}

The spin drift-diffusion equation (\ref{general}), together with the local
charge
neutrality constraint Eq. (\ref{neutral}),
dramatically alters the spin transport behavior in semiconductors
from that expected from Eq. (\ref{deg}).
The general solution to Eq. (\ref{deg}) (restricting variation to the
$x$-direction) is
\begin{equation}
\mu_{\uparrow}-\mu_{\downarrow}= A_1\exp(-x/L) +A_2 \exp(x/L),
\end{equation}
where $A_1$ and $A_2$ are constants.
In contrast the general form of solution to Eq. (\ref{nondeg})
is
\begin{equation}
n_{\uparrow}-n_{\downarrow}=A_1 \exp(-x/L_1) +A_2 \exp (- x/L_2),
\end{equation}
where $\lambda_1=1/L_1$ and $\lambda_2 =1/L_2$ are the roots of
the quadratic equation,
\begin{equation}
\lambda^2-\lambda \nu E/D
-1/L^2=0.
\end{equation}

To understand the physical consequence of the electric field on the spin
diffusion,
we suppose that a continuous spin imbalance is injected at $x=0$,
$(n_{\uparrow}-n_{\downarrow})|_0$,
and the electric
field is along the
$-x$ direction.
The spin polarization will gradually
decay in size as the distance
from the point of injection increases and eventually go to zero at $\pm
\infty$.
The distribution of the spin polarization then can be
described by
\begin{subequations}
\begin{equation}
n_{\uparrow}-n_{\downarrow} = (n_{\uparrow}-n_{\downarrow})|_0\exp
(-x/L_d),~x>0,
\label{d-stream}
\end{equation}
\begin{equation}
n_{\uparrow}-n_{\downarrow}= (n_{\uparrow}-n_{\downarrow})|_0\exp
(x/L_u),~x<0,
\label{u-stream}
\end{equation}
\end{subequations}
where we define two quantities $L_d$ and $L_u$ as the down-stream
and up-stream spin diffusion lengths, respectively,
\begin{subequations}
\begin{equation}
L_d=\bigg[-\frac{|eE|}{2}\frac{\nu}{eD}+
\sqrt{\Big(\frac{|eE|}{2}\frac{\nu}{e D}\Big)^2+\frac{1}{L^2}}~\bigg]^{-1},
\label{d-length_g}
\end{equation}
\begin{equation}
L_u=\bigg[\frac{ |eE|}{2}\frac{\nu}{eD}+
\sqrt{\Big(\frac{|eE|}{2}\frac{\nu}{eD}\Big)^2+\frac{1}{L^2}}~\bigg]^{-1},
\label{u-length_g}
\end{equation}
\end{subequations}
and $L_u L_d =L^2$. Here $\nu/eD$ in 3D and  2D
systems
can be evaluated via Eqs. (\ref{3d}) and (\ref{2d}), respectively.
For nondegenerate semiconductors, Eqs. (\ref{d-length_g}) and
(\ref{u-length_g})
reduce to
\begin{subequations}
\begin{equation}
L_d=\bigg[-\frac{|eE|}{2k_B T}+ \sqrt{\Big(\frac{e E}{2 k_B
T}\Big)^2+\frac{1}{L^2}}~\bigg]^{-1},
\label{d-length}
\end{equation}
\begin{equation}
L_u=\bigg[\frac{|eE|}{2k_B T}+ \sqrt{\Big(\frac{e E}{2k_B
T}\Big)^2+\frac{1}{L^2}}~\bigg]^{-1}.
\end{equation}
\end{subequations}
Equation (\ref{d-length}) was reported in Ref.
\onlinecite{aronov}.

In the absence of the field,
the down-stream and up-stream lengths are equal to the intrinsic
diffusion length $L$. With increasing field the down-stream
diffusion length $L_d$ increases, whereas the up-stream diffusion length
$L_u$ decreases. It was also shown in Ref. \onlinecite{aronov} that
spin transport distance in semiconductors can be increased by an electric
field (down-stream
diffusion length), but the up-stream diffusion length was not discussed.
A high-field regime for spin transport in semiconductors
can be defined by $E > E_c$, where $eE_c/k_BT =1/L$.
In this regime, $L_u$ and $L_d$ deviate from $L$
considerably and the spin diffusion behavior
is qualitatively different from that in low fields.
We emphasize that since $L$ is large in semiconductors, this
regime
is not beyond realistic fields where most spintronic devices operate.
For a typical spin diffusion length in semiconductors, $L=2$
$\mu$m,\cite{length}
$E_c=125$ V/cm
at $T=300$ K and $E_c=1.25$ V/cm at $T=3$ K.

The physics of the field effects on the spin diffusion becomes
clearer at the strong-field limit, where
$|eE|/k_B T \gg 1/L$. In this limit,
the electrons move
with drift velocity $|E| \nu_e$ and so does the spin polarization.
$L_d$ is simply the distance over which the carriers move within
the spin life time $\tau_S$,\cite{book,aronov}
\begin{subequations}
\begin{equation}
L_d  \simeq \frac{|eE|}{k_B T} L^2 =
 \frac{e|E|}{k_B T} D\tau_S =\nu_e|E| \tau_S.
\end{equation}
For the up-stream diffusion length $L_u$ at this limit,
\begin{equation}
L_u \simeq k_B T /|eE|,
\end{equation}
\end{subequations}
which simply corresponds to a Boltzmann
distribution of electrons in a retarding field.\cite{book}

\subsection{Carrier densities versus electrochemical potentials}

In the literature of spin transport in metals, the spin polarization
is usually described by the splitting of
electrochemical potentials for up-spin and down-spin electrons.
For nondegenerate semiconductors,
the density difference between up-spin and down-spin electrons
is a natural way to characterize the spin polarization. It is therefore
useful to establish the connection between these two quantities.

The electrochemical potentials for up-spin and down-spin
electrons in a semiconductor are
related to their densities via
\begin{equation}
n_{\uparrow(\downarrow)}=n^0_{\uparrow(\downarrow)}\bigg[\exp\bigg(\frac{\mu_{\uparrow(\downarrow)}-\mu_0}{k_B
T}
\bigg)-1\bigg],
\end{equation}
where $\mu_0$ is the value that the  electrochemical potential would have
without spin polarization,
\begin{equation}
{\bf \nabla}\mu_0=(e/\sigma_s) {\bf J},
\end{equation}
where $\sigma_s$ is the electrical conductivity of the semiconductor and
{\bf J} is the total electrical current. In a doped
semiconductor with a homogeneous carrier concentration, the electrochemical
potential $\mu_0$ at postion {\bf x} is given by
\begin{equation}
\mu_0=(e/\sigma_s){\bf J}\cdot {\bf x} -B=e {\bf E}\cdot {\bf x} -B,
\end{equation}
where $B$ is a constant. Thus the electrochemical potentials for individual
spins are
\begin{equation}
\mu_{\uparrow(\downarrow)}=k_B T \ln
\bigg(1+\frac{n_{\uparrow(\downarrow)}}{n^0_{\uparrow(\downarrow)}}\bigg)
+ e {\bf E}\cdot {\bf x} -B.
\end{equation}
The electrochemical potential splitting, $\mu_{\uparrow}-\mu_{\downarrow}$,
and the density difference, $n_{\uparrow}-n_{\downarrow}$, between up-spin
and down-spin
electrons then are related via
\begin{equation}
\mu_{\uparrow}-\mu_{\downarrow}=k_B T
\ln\bigg[\frac{1+(n_{\uparrow}-n_{\downarrow})/2n^0_{\uparrow}}
{1-(n_{\uparrow}-n_{\downarrow})/2n^0_{\downarrow}}\bigg].
\end{equation}
Therefore it is advantageous to use $n_{\uparrow}-n_{\downarrow}$ instead of
$\mu_{\uparrow}-\mu_{\downarrow}$
to describe spin transport in semiconductors, for the spin drift-diffusion
equation is linear in terms of the former,
but would be nonlinear in terms of the latter.

When $\mu_{\uparrow}-\mu_{\downarrow} \ll k_B T$,
\begin{equation}
(n_{\uparrow}-n_{\downarrow})\Big(\frac{1}{2n^0_{\uparrow}}+\frac{1}{2n^0_{\downarrow}}\Big)=
\frac{\mu_{\uparrow}-\mu_{\downarrow}}{ k_B T},
\end{equation}
and we have the drift-diffusion equation for  the electrochemical potential
splitting,
\begin{equation}
\nabla^2(\mu_{\uparrow}-\mu_{\downarrow})+\frac{e{\bf E}}{k_B T}
\cdot{\bf\nabla}(\mu_{\uparrow}-\mu_{\downarrow})
-\frac{\mu_{\uparrow}-\mu_{\downarrow}}{L^2}=0.
\label{nondeg_mu}
\end{equation}
In this linear differential equation for $\mu_{\uparrow}-\mu_{\downarrow}$,
the electric
field still plays a central role and there are two distinct diffusion
lengths, i.e.,
the down-stream ($L_d$) and the up-stream ($L_u$) diffusion lengths, for
$\mu_{\uparrow}-\mu_{\downarrow}$.
Thus spin transport predicted by Eq. (\ref{nondeg_mu})
would be still
qualitatively different from that expected from Eq. (\ref{deg}).

\subsection{Current versus density spin polarization}

There exist two definitions in literature to
characterize spin polarization in nonmagnetic semiconductors. One definition
uses the density difference between up-spin and down-spin
electrons,
\begin{equation}
P(x)\equiv \frac{n_{\uparrow}-n_{\downarrow}}{n^0},
\end{equation}
where $n_0=2n_{\uparrow(\downarrow)}$ is the total electron density.
The other uses the current difference between up-spin and down-spin
electrons,
\begin{equation}
\alpha(x)\equiv
\frac{j_{\uparrow}-j_{\downarrow}}{j_{\uparrow}+j_{\downarrow}}.
\end{equation}

Generally speaking, these two spin polarization are different, although they
are related.
To find the
relationship between these two polarizations in a homogeneous
nonmagnetic semiconductor, we note that
\begin{equation}
j_{\uparrow}-j_{\downarrow}
=e(n_{\uparrow}-n_{\downarrow})\nu E +e D
\frac{d(n_{\uparrow}-n_{\downarrow})}{dx}.
\end{equation}
By using  the local charge neutrality condition Eq. (\ref{neutral}), we 
obtain
\begin{equation}
\alpha(x)=P(x)+\frac{D}{\nu E}\frac{dP}{dx}.
\end{equation}
For a steady spin imbalance injected at $x=0$, as discussed in Sec. II.B,
according to the general
solution of  Eqs. (\ref{d-stream}) and (\ref{u-stream}),
\begin{eqnarray}
\frac{d(n_{\uparrow}-n_{\downarrow})}{dx}&=&-\frac{1}{L_d}(n_{\uparrow}-n_{\downarrow}),~~~~~~~~x>0,\\
%\end{equation}
%\begin{equation}
\frac{d(n_{\uparrow}-n_{\downarrow})}{dx}&=&\frac{1}{L_u}(n_{\uparrow}-n_{\downarrow}),~~~~~~~~x<0,
\end{eqnarray}
and the relation between the spin polarization of current $\alpha(x)$
and the spin polarization of density $P(x)$
can be written as
\begin{equation}
\alpha(x)=P(x) \bigg[1-\Big(\frac{\nu}{e D}\Big)^{-1}\frac{1}{e E L_d}\bigg]
\label{ap-d_g}
\end{equation}
for $x>0$, and
\begin{equation}
\alpha(x)=P(x) \bigg[1+\Big(\frac{\nu}{e D}\Big)^{-1}\frac{1}{e E L_u}\bigg]
\label{ap-u_g}
\end{equation}
for $x<0$. $\nu/eD$ is Eqs. (\ref{ap-d_g}) and (\ref{ap-u_g}) can be
calculated using Eqs. (\ref{3d}) and (\ref{2d}) for 3D and 2D systems
with different doping concentrations and temperatures. In the nondegenerate
limit, Eq. (\ref{ap-d_g}) reduces to
\begin{equation}
\alpha(x)=P(x) \bigg(1-\frac{k_B T}{e E L_d}\bigg),
\label{ap-d}
\end{equation}
which is equivalent to Eq. (5) in Ref. \onlinecite{aronov}, where
the authors studied magnetization ($P$) in the presence of current
with a {\em given} spin polarization ($\alpha$) in semiconductors.
In the meantime Eq. (\ref{ap-u_g}) reduces to
\begin{equation}
\alpha(x)=P(x) \bigg(1+\frac{k_B T}{e E L_u}\bigg).
\label{ap-u}
\end{equation}
Thus in semiconductors
$\alpha(x)$ is
proportional to  $P(x)$,
and the ratio between them  depends on the electric field and its direction.

\section{Field enhanced spin injection in FM/NS structures}

We first  consider a simple one-dimensional spin injection structure
to elucidate the underlying physics of
electric-field enhanced spin injection.
This injection structure comprises a semi-infinite degenerate ferromagnet
($x <0$) and a semi-infinite
nonmagnetic nondegenerate semiconductor ($x>0$).
Electrons are injected from the magnet into the semiconductor, and
therefore,
the electric field is antiparallel to the $x$-axis.
In the ferromagnet
the electrochemical potentials for individual spins
satisfy Eq. (\ref{e_potential}), which has the following general solution:
\begin{equation}
\frac{1}{eJ}\left ( \begin{array}{c} \mu_{\uparrow}\\
\mu_{\downarrow}
\end{array} \right )
=\frac{x}{\sigma^f_{\uparrow}+\sigma^f_{\downarrow}} \left (
\begin{array}{c} 1\\ 1
\end{array} \right )+ C  \left
( \begin{array}{c}1/\sigma^f_{\uparrow}\\ -1/\sigma^f_{\downarrow}
\end{array} \right )e^{x/L^{(f)}},
\end{equation}
where $\sigma^f_{\uparrow(\downarrow)}$ is the up-spin (down-spin)
electrical conductivity of the ferromagnet, and
$J$ is the total electron current, which is a constant throughout the
structure
in steady state. We use $L^{(f)}$ and $L^{(s)}$ to denote the intrinsic
spin diffusion length in the ferromagnet and in the semiconductor,
respectively.

In the semiconductor, up-spin and down-spin electron densties satisfy the
spin drift-diffusion equation for nondegenerate systems,  Eq.
(\ref{nondeg}),
as well as the local charge neutrality
condition, Eq. (\ref{neutral}). The general
solution can be written as
\begin{equation}
n_{\uparrow(\downarrow)}=+(-) A \exp (-x/L_d),
\label{solution_n}
\end{equation}
and accordingly the electrochemical potentials for individual spins are
\begin{equation}
\mu_{\uparrow(\downarrow)}=k_B T \ln \Big[1+(-)\frac{2A
e^{-x/L_d}}{n_0}\Big]+e Ex -B.
\end{equation}
Here $n_0$ is total electron density.

In general, a Schottky barrier will form between the magnet and the
semiconductor when the
two materials are placed together. Since  the charge neutrality condition
would be severely violated
in the depletion region of a Schottky barrier, a wide depletion region is
undesirable
for spin transport and spin coherence. Specifically the presence of holes
dramatically shortens the electron spin coherence time.
We consider instead structures with a very thin interfacial barrier between
the magnet
and the
semiconductor. A spin-selective interfacial barrier was examined
in Refs. \onlinecite{rashba,smith,fert,spin_book} as a way to
circumvent the resistance mismatch obstacle for spin injection
from a ferromagnetic metal into a semiconductor. If there is no
spin-flip scattering at the interface, the current for an
individual spin is continuous across the interface and is related
to the spin-dependent electrochemical potential change across the
interface via Ohmic's law, giving rise to the following boundary
conditions:
\begin{subequations}
\begin{eqnarray}
j_{\uparrow}(0^-)&=&G_{\uparrow}[\mu_{\uparrow}(0^+)
-\mu_{\uparrow}(0^-)],\\
j_{\downarrow}(0^-)&=&G_{\downarrow}[\mu_{\downarrow}(0^+)
-\mu_{\downarrow}(0^-)],\\
j_{\uparrow}(0^-)&-&j_{\downarrow}(0^-) =
j_{\uparrow}(0^+)-j_{\downarrow}(0^+),
\end{eqnarray}
\end{subequations}
where $G_{\uparrow(\downarrow)}$ is the interfacial conductance
for up-spin (down-spin) electrons and the current of individual spins
$j_{\uparrow(\downarrow)}$ can be calculated via
$ej_{\uparrow(\downarrow)}=\sigma_{\uparrow(\downarrow)}d\mu_{\uparrow(\downarrow)}/dx$.
These three equations completely determine
the three unknown coefficients $A$, $B$, $C$ in Eqs. (3.1)-(3.3).

The solution of $n_{\uparrow(\downarrow)}$ in Eq. (\ref{solution_n}) and
the relation of Eq. (\ref{ap-d}) indicate that in the semiconductor
$\alpha (x)= \alpha_0 e^{-x/L_d}$, where $\alpha_0$ is
the spin injection efficiency at the interface.
We obtain an equation for $\alpha_0$,
\begin{widetext}
\begin{equation}
\frac{G^{-1}_{\uparrow}-G^{-1}_{\downarrow}}{2}
+\frac{p_f(G^{-1}_{\uparrow}+G^{-1}_{\downarrow})}{2}
+(\alpha_0-p_f)\bigg[\frac{G^{-1}_{\uparrow}+G^{-1}_{\downarrow}}{2}+
\frac{2L^{(f)}}{(1-p^2_f)\sigma_f}\bigg]
=\frac{k_B T}{eE
\sigma_s} \ln \frac{-k_BT/e E L_u +\alpha_0}{-k_BT/e E L_u
-\alpha_0},
\end{equation}
\end{widetext}
where
$\sigma_f=\sigma^f_{\uparrow}+\sigma^f_{\downarrow}$ is the conductivity of
the ferromagnet,
$\sigma_s=n_0 e \nu$ the conductivity of the semiconductor, and
$p_f=(\sigma^f_{\uparrow}-\sigma^f_{\downarrow})/\sigma_f$ the spin
polarization in
the ferromagnet.

We solve this equation and plot the spin injection of current
$\alpha_0$ as a function of the electric field in Fig. 2.
We see that the electric field can substantially enhance the spin injection
efficiency in FM/NS structures.
We note that spin injection enhancement from a
spin-selective interfacial barrier between the ferromagnet
and the semiconductor, which has been identified in the low-field
regime,\cite{rashba,smith,fert} becomes more pronounced in the high-field
regime.
\begin{figure}
\vspace{10pt}\includegraphics[width=7cm]{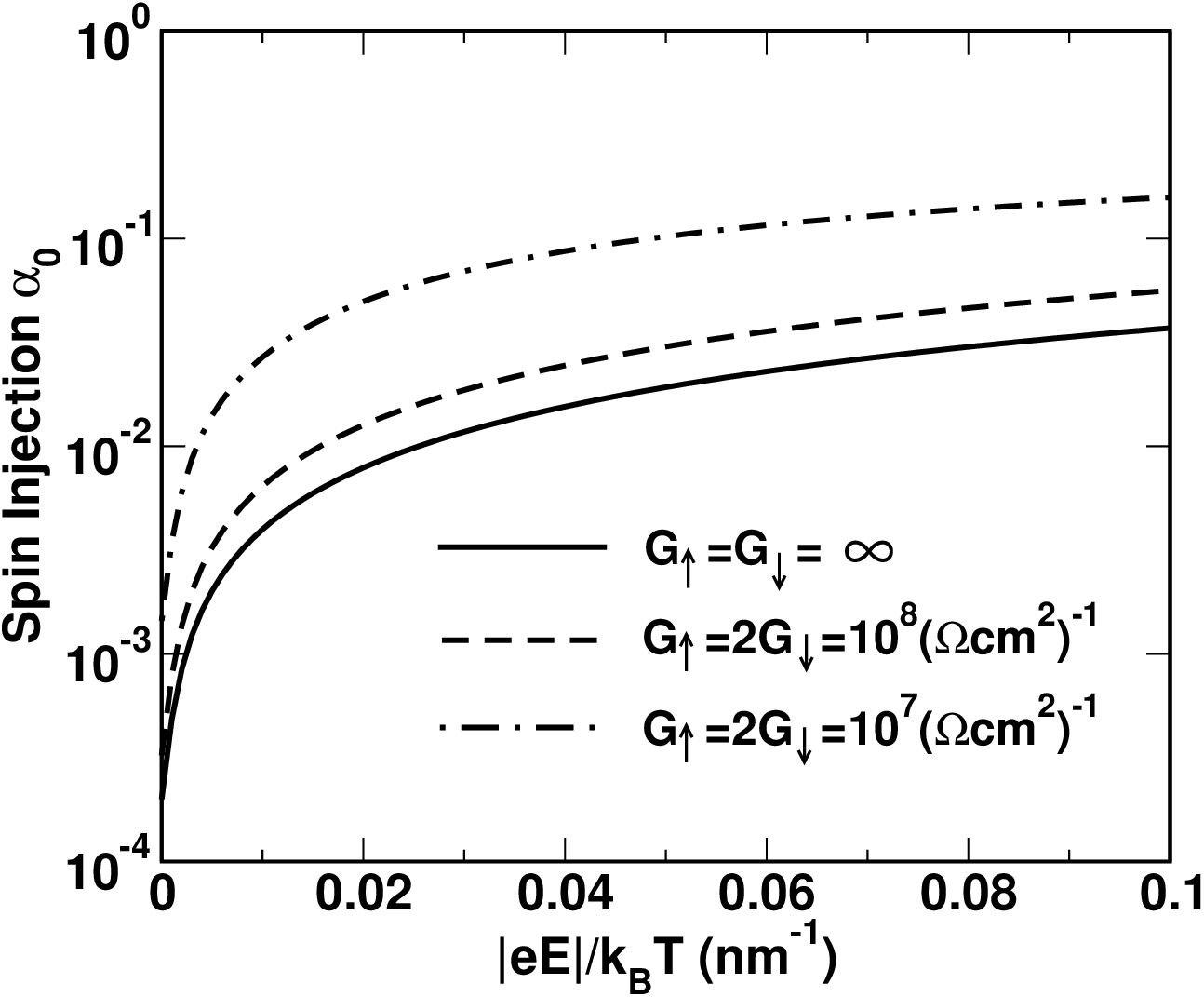}
\caption{Spin injection efficiency $\alpha_0$ as a function of electric
field. The interfacial conductance for up-spin electrons is twice larger
than that for down-spin electrons, $G_{\uparrow}=2G_{\downarrow}$.
Solid, dashed, and dot-dashed lines correspond to $G_{\uparrow}=\infty$
(transparent interface),
10$^8$, and 10$^7~(\Omega$ cm$^2)^{-1}$, respectively.
Other parameters are $p_f=0.5$, $L^{(f)}=60$ nm,
$L^{(s)}=2$ $\mu$m, and $\sigma_f=100~\sigma_s=10^3$ ($\Omega$ cm)$^{-1}$.}
\vspace{10pt}
\end{figure}

In the small spin polarization limit, $ n_{\uparrow(\downarrow)}/n_0 \ll 1$,
$\alpha(x)$ can be expressed in an explicit form,
\begin{eqnarray}
\alpha(x) &=&\bigg[\frac{L^{(f)}}{(1-p^2_f)\sigma_f}
+\frac{L_u}{\sigma_s}+\frac{G_{\uparrow}+G_{\downarrow}}
{4G_{\uparrow}G_{\downarrow}}\bigg]^{-1}\nonumber\\
&\times&\bigg[\frac{p_f
L^{(f)}}{(1-p^2_f)\sigma_f}+\frac{G_{\uparrow}-G_{\downarrow}}{4G_{\uparrow}G_{\downarrow}}\bigg]
e^{-x/L_d}.
\label{ex_fn}
\end{eqnarray}
This expression clearly shows that the electric field and the spin-selective
interfacial resistance
both enhance spin injection, but in  different ways.
The electric-field
effects on spin injection can be
described in terms of
the two field-induced diffusion lengths. Both diffusion lengths
affect spin injection favorably:
The up-stream length $L_u$ controls the relevant resistance in the
semiconductor, which
determines the spin injection efficiency. With increasing field this
effective resistance, $L_u/\sigma_s$, becomes smaller, and accordingly
the spin injection efficiency is enhanced.
The transport distance of the injected spin polarization in the
semiconductor,
however, is controlled by the down-stream length $L_d$. As
the field increases, this distance becomes longer. On the other hand,
the
spin-selective interfacial barrier provides another spin polarization source
besides the spin-aligned
electrons in the ferromagnet and acts as a spin filter which permits
electrons
with a particular spin to pass through the interface, and therefore enhances
spin injection. Moreover, the spin-selective barrier has no effect on the
transport distance of the injected
spin polarization in the semiconductor.

We now contrast Eq. (\ref{ex_fn}) with that obtained by previous
calculations \cite{schmidt,rashba,smith,fert} based on Eq. (\ref{deg}).
The spin injection
%\begin{widetext}
\begin{eqnarray}
\alpha(x)&=&\Big[\frac{L^{(f)}}{(1-p^2_f)\sigma_f}+
\frac{L^{(s)}}{\sigma_s}+\frac{G_{\uparrow}+G_{\downarrow}}
{4G_{\uparrow}G_{\downarrow}}\Big]^{-1}\nonumber\\
&\times&\Big[\frac{p_f
L^{(f)}}{(1-p^2_f)\sigma_f}+\frac{G_{\uparrow}-G_{\downarrow}}{4G_{\uparrow}G_{\downarrow}}\Big]e^{-x/L^{(s)}}
\label{zero_e}
\end{eqnarray}
%\end{widetext}
is given by the zero-field result of Eq. (\ref{ex_fn}). For a transparent
interface
with $G^{-1}_{\uparrow}=G^{-1}_{\downarrow}=0$, the
effective resistance in the magnet, $L^{(f)}/\sigma_f$, is much less than
its
counterpart in the semiconductor, $L^{(s)}/\sigma_s$ (as
$L^{(f)} \ll L^{(s)}$ and $\sigma_f \gg \sigma_s$).
Thus Eq. (\ref{zero_e}) suggests that
this resistance mismatch makes
it virtually impossible to realize an appreciable spin injection
from a ferromagnetic metal to a semiconductor without spin-selective
interfacial barrier.
However, the more general description of the spin transport in
semiconductors
indicates
that
the effective semiconductor resistance to be compared with
$L^{(f)}/\sigma_f$
should be $L_u/\sigma_s$ rather than $L^{(s)}/\sigma_s$. Since $L_u$ can be
smaller than $L^{(s)}$ by orders of magnitude in the high-field regime,
this ``conductivity mismatch'' obstacle may be overcome with the help
of strong electric fields, or equivalently, large injection
currents. We note that although it has been realized that spin injection can
be enhanced
by increasing the total injection current,\cite{smith} the treatment there
used
Eq. (\ref{deg}) to describe spin transport in nondegenerate semiconductors,
where
the electric-field effects were not taken into account. Thus the physics of
the
field-dependent spin transport was not captured and the treatment was
incomplete.

In the presence of the interfacial barrier, the relative importance of the
two mechanisms
for the spin injection efficiency enhancement, electric field and
spin-selective interfacial barrier,
depends on the relative magnitude of  $R_f=
L^{(f)}/(1-p^2_f)\sigma_f$,  $R_s \equiv L^{(s)}/\sigma_s$,
and $R_i \equiv (G_{\uparrow}+G_{\downarrow})/
4G_{\uparrow}G_{\downarrow}$, the resistances of the ferromagnet, the
semiconductor,
and the interface,
which may vary from system to system.
For systems with $R_i \gg R_s \gg R_f$, the spin-selective interfacial
barrier dominates in enhancing the spin injection efficiency. On the other
hand,
for systems with $R_s \gg R_i \gg R_f$ or $R_s \gg R_f \gg R_i$,
the electric field can enhance spin injection efficiently.
The transport distance of injected spin polarization in the semiconductor,
however,
can only be enhanced by the electric field, even in systems with $R_i \gg
R_s \gg R_f$.
Furthermore, the two mechanisms have different temperature and field
dependences. The field-enhanced spin injection efficiency
increases with the electric field (current) and decreases with the
temperature.
The interface-enhanced spin injection likely depends on neither the field
nor the temperature.

To quantitatively demonstrate the electric-field effects on spin injection,
we choose the realistic parameters of a spin injection device as follows:
$p_f=0.5$ and $L^{(f)}=60$ nm (as in Co),\cite{material} $L^{(s)}=2$ $\mu$m
(as in GaAs).\cite{length}
First we examine the structures without an interfacial barrier.
For a ferromagmetic metal/semiconductor structure, e.g., Co/GaAs,
$\sigma_f \simeq10^4~\sigma_s$, the spin injection efficiency
increases from $2\times 10^{-6}$ at zero field to 2\% at
$|eE|/k_BT=5$ nm$^{-1}$. For a ferromagnetic semiconductor
with $p_f \sim 0.5$ and
$\sigma_f \sim 100 ~\sigma_s$,
the spin injection efficiency increases from 0.02\% at zero field
to 2\% at $|eE|/k_BT=0.05$ nm$^{-1}$, which
corresponds to $|E|=125$ V/cm, or $|J|=1250$ A/cm$^2$
for a typical
semiconductor conductivity $\sigma_s=10$ ($\Omega$ cm)$^{-1}$, at $T=3$ K.

Suppose there existed a spin-selective interfacial barrier between
the magnet and the semiconductor with $G_{\uparrow}=2G_{\downarrow}=10^7$
($\Omega$ cm$^2$)$^{-1}$.
For the Co/GaAs structure, the spin injection efficiency would
be 0.1\% at zero field (much greater than $2\times 10^{-6}$
in a similar injection structure without a spin-selective interfacial
barrier),
which can be further enhanced to 10\% at $|eE|/k_BT=0.06$ nm$^{-1}$, or
$E=15$ kV/cm at $T=300$ K.
For the spin injection structure from the ferromagnetic semiconductor,
the spin injection efficiency is 0.1\% at zero field (also considerably
greater than 0.02\%
in a similar structure without a spin-selective interfacial barrier),
which can be further increased to 10\% at $|eE|/k_BT=0.05$ nm$^{-1}$.
Thus the combination of electric field and spin-selective interfacial
barrier may help
explain the large spin injection
percentages from ZnMnSe to ZnSe,\cite{molenkamp,jonker1} from
GaMnAs to GaAs,\cite{ohno}
from Fe to GaAs,\cite{ploog,jonker2,motsnyi} as well as the
dramatic increase in spin injection with current in Ref.
\onlinecite{jonker2}.

Figure 3 shows  spin polarization of current ($\alpha_0$)  and  spin
polarization of
density ($P_0$) at the interface as
a function of electric field. According to Eq. (\ref{ap-d}),
$P_0=\alpha_0 (|eE|L_u/k_BT)$. We can see that
in the low-field regime, the density polarization in the semiconductor is
much smaller than
the current polarization; whereas in the high-field regime, the two
polarizations
become equal. Thus
the difference between high-field injection and low-field injection is that
a strong electric field in high-field injection
results in a macroscopic density difference between up-spin and down-spin
electrons; whereas
in low-field injection the carrier densities of up-spin and down spin
electrons
remain the same.
\begin{figure}
\vspace{10pt}
\includegraphics[width=7cm]{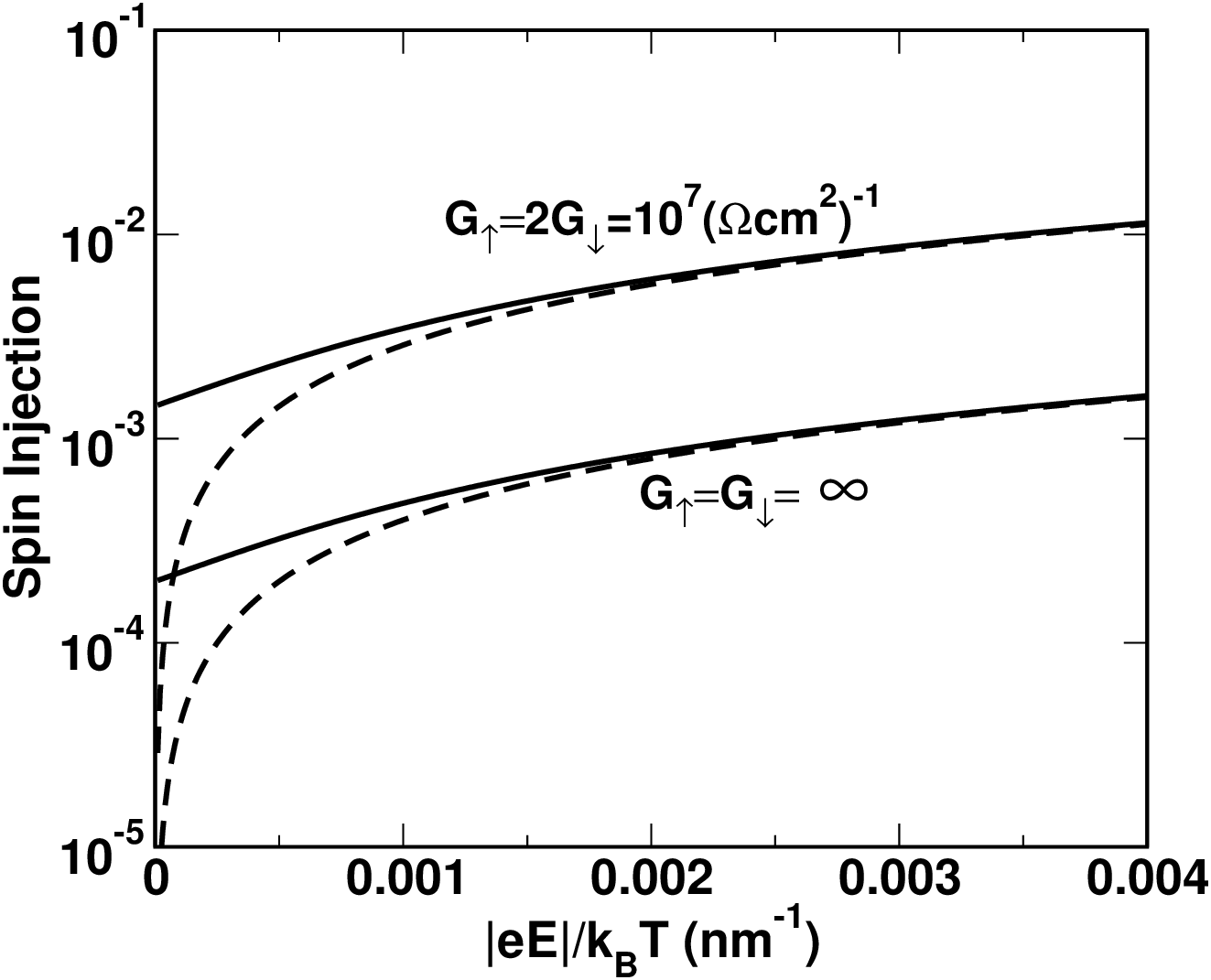}
\caption{Spin injection as a function of electric
field.
Solid and dashed lines describe spin polarization of current $\alpha_0$
and spin polarization of density $P_0$ at the interface, respectively.
The lower curves are for transparent interface structure and
the upper ones are for structure with a spin selective interfacial barrier,
$G_{\uparrow}=2G_{\downarrow}=10^7~(\Omega$ cm$^2)^{-1}$.
Other parameters are the same as in Fig. 2. }
\vspace{10pt}
\end{figure}

Another important quantity is the boundary resistance $R_b$,
which can be used to analyze how easy a current can convert
its spin while flowing from the magnet into the semiconductor,
\begin{equation}
R_b=\frac{\mu_0(0^+)-\mu_0(0^-)}{eJ},
\end{equation}
which can be expressed in terms of the spin injection efficiency and
the resistances of the magnet, the semiconductor, and the interface.
\begin{eqnarray}
R_b&=&\frac{-B}{eJ}=(p_f-\alpha_0)\bigg[ \frac{p_f
L^{(f)}}{(1-p^2_f)\sigma_f}+
\frac{G_{\uparrow}-G_{\downarrow}}{4G_{\uparrow}G_{\downarrow}}\bigg]\nonumber\\
&+&\frac{G_{\uparrow}+G_{\downarrow}}{4G_{\uparrow}G_{\downarrow}}
-p_f\frac{G_{\uparrow}-G_{\downarrow}}{4G_{\uparrow}G_{\downarrow}}.
\end{eqnarray}
We plot $R_b$ in Fig. 4 as a function of electric field.
With increasing field, the boundary resistance decreases, indicating
that the field helps
current conversion from a higher polarization (in the magnet)
to a lower polarization (in the semiconductor).
For structures with a transparent interface, in the small polarization limit
[$n_{\uparrow(\downarrow)}/n_0 \ll 1$],
\begin{equation}
R_b=\bigg[\frac{L^{(f)}}{\sigma_f}+\frac{L_u}{\sigma_s}(1-p^2_f)\bigg]^{-1}
p^2_f\frac{L^{(f)}}{\sigma_f}\frac{L_u}{\sigma_s},
\end{equation}
which clearly shows that the boundary resistance is determined by the
up-stream diffusion length
$L_u$ rather than $L^{(s)}$ and should be a function of electric field.
\begin{figure}
\vspace{10pt}
\includegraphics[width=7cm]{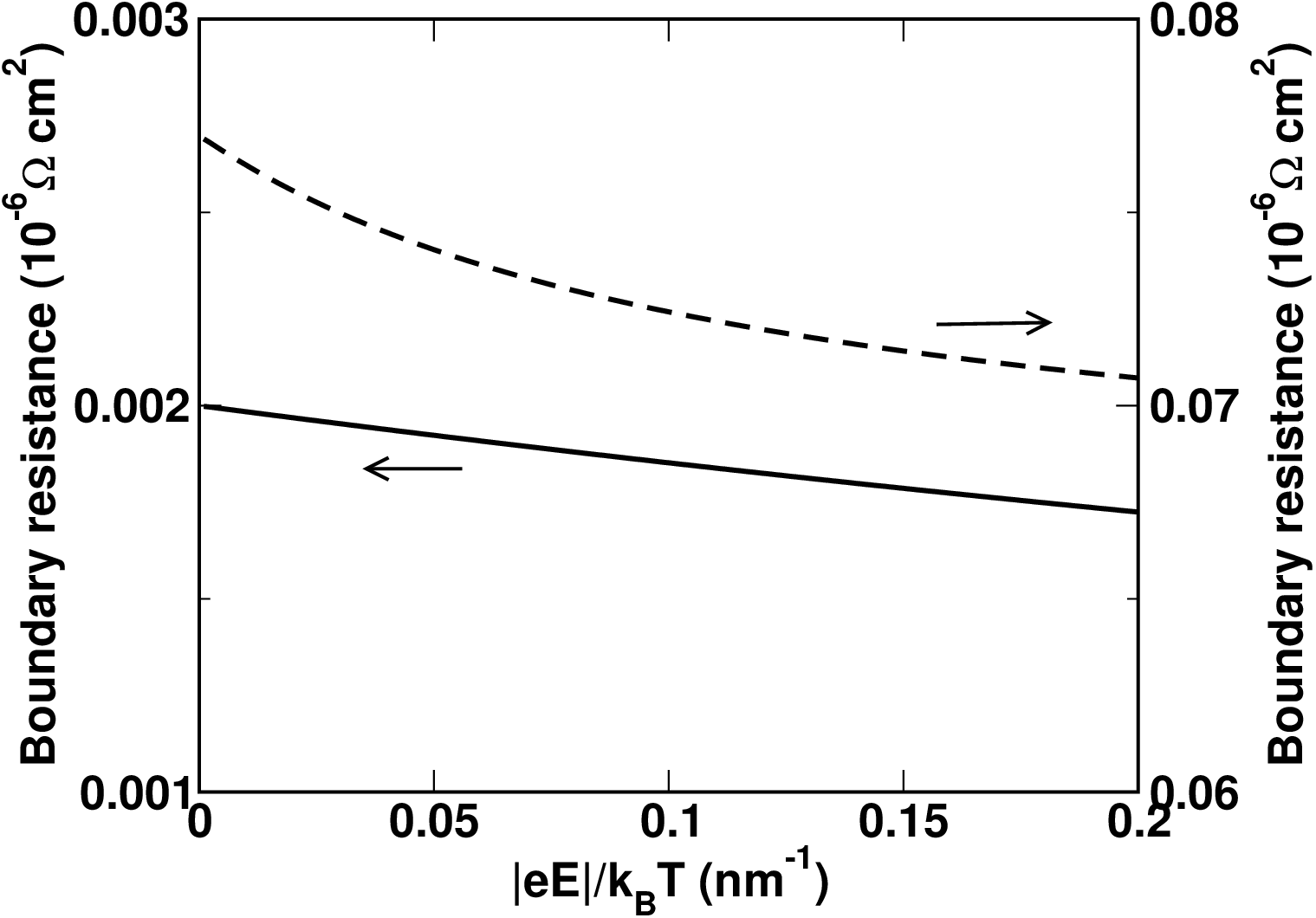}
\caption{Boundary resistance  $R_b$ as a function of electric
field.
Solid line is for structure with transparent interface. Dashed line is
for structure with a spin-selective interfacial barrier,
$G_{\uparrow}=2G_{\downarrow}=10^7~(\Omega$ cm$^2)^{-1}$.
Other parameters are the same as in Fig. 2. }
\vspace{10pt}
\end{figure}

\section{Spin injection in FM/NS/FM structures}

In this section we consider a sandwiched structure
that comprises a semi-infinite ferromagnet ($x <0$),
a nonmagnetic semiconductor with width $x_0$, and a semi-infinite
ferromagnet
($x>x_0$). $x_0$ is assumed to be much shorter than the semiconductor
intrinsic
spin diffusion length $L^{(s)}$, as in most spintronic devices.
The two ferromagnets are otherwise identical with
possible different orientations of magnetization (parallel and
anti-parallel).
Electrons are injected from the left magnet to the semiconductor and
the electric field is antiparallel to the $x$-axis.
The spin transport in the magnets is described by Eq. (\ref{e_potential})
with the following
general solutions:
\begin{equation}
\frac{1}{eJ}\left ( \begin{array}{c} \mu_{\uparrow}\\
\mu_{\downarrow}
\end{array} \right )
=\frac{x}{\sigma^L_{\uparrow}+\sigma^L_{\downarrow}} \left (
\begin{array}{c} 1\\ 1
\end{array} \right )+ C_L  \left
( \begin{array}{c}1/\sigma^L_{\uparrow}\\ -1/\sigma^L_{\downarrow}
\end{array} \right )e^{x/L^{(f)}}
\end{equation}
in the left magnet ($x<0$), and
\begin{eqnarray}
\frac{1}{eJ}\left ( \begin{array}{c} \mu_{\uparrow}\\
\mu_{\downarrow}
\end{array} \right )
&=&\Big(\frac{x}{\sigma^R_{\uparrow}+\sigma^R_{\downarrow}}+B_1\Big) \left (
\begin{array}{c} 1\\ 1
\end{array} \right )\nonumber\\
&+& C_R  \left
( \begin{array}{c}1/\sigma^R_{\uparrow}\\ -1/\sigma^R_{\downarrow}
\end{array} \right )e^{-(x-x_0)/L^{(f)}}
\end{eqnarray}
in the right magnet ($x>x_0$).
Here we use superscript or subscript $L$ ($R$) to label the quantities in
the left (right)
ferromagnet.

The spin-dependent electron densities in the semiconductor satisfy Eqs.
(\ref{nondeg}) and (\ref{neutral}),
and the general solutions at $0<x<x_0$ can be written as
\begin{equation}
n_{\uparrow(\downarrow)}=+(-) \Big[ A_0 e^{-x/L_d} +A_1 e^{(x-x_0)/L_u}
\Big],
\end{equation}
and therefore the electrochemical potentials for individual spins are
\begin{eqnarray}
\mu_{\uparrow(\downarrow)}&=&k_B T \ln \bigg[1+(-)\frac{2A_0e^{-x/L_d}+2A_1
e^{(x-x_0)/L_u}}{n_0}\bigg]\nonumber\\
&+&e Ex -B_0.
\end{eqnarray}

The six unknowns in the above solutions, $A_0$, $A_1$, $B_0$, $B_1$, $C_L$,
and $C_R$,
are determined by the following six independent boundary conditions:
\begin{subequations}
\begin{eqnarray}
j_{\uparrow}(0^-)&=&G_{\uparrow}[\mu_{\uparrow}(0^+)-\mu_{\downarrow}(0^-)],\\
j_{\downarrow}(0^-)&=&G_{\downarrow}[\mu_{\downarrow}(0^+)-\mu_{\downarrow}(0^-)],\\
j_{\uparrow}(0^-)&-&j_{\downarrow}(0^-) =
j_{\uparrow}(0^+)-j_{\downarrow}(0^+),\\
j_{\uparrow}(x_0^+)&=&{\tilde
G}_{\uparrow}[\mu_{\uparrow}(x_0^+)-\mu_{\downarrow}(x_0^-)],\\
j_{\downarrow}(x_0^+)&=&{\tilde
G}_{\downarrow}[\mu_{\downarrow}(x_0^+)-\mu_{\downarrow}(x_0^-)],\\
j_{\uparrow}(x_0^-)&-&j_{\downarrow}(x_0^-) =
j_{\uparrow}(x_0^+)-j_{\downarrow}(x_0^+),
\end{eqnarray}
\end{subequations}
where $G_{\uparrow(\downarrow)}$ and $\tilde{G}_{\uparrow(\downarrow)}$
are the spin-dependent conductances at the left and the right interfaces,
respectively.

We find the two equations for the spin polarizations of current at
interfaces, $\alpha(0)$ and
$\alpha(x_0)$, by matching the above boundary conditions, Eqs. (4.5a-f),
\begin{widetext}
\begin{equation}
\left\{ \begin{array}{c}
{\displaystyle
\frac{p_L(G_{\uparrow}+G_{\downarrow})}{2G_{\uparrow}G_{\downarrow}}
-\frac{G_{\uparrow}-G_{\downarrow}}{2G_{\uparrow}G_{\downarrow}}
+\Big[\alpha(0)-p_L\Big]\bigg[\frac{G_{\uparrow}+G_{\downarrow}}{2G_{\uparrow}G_{\downarrow}}+
\frac{2L^{(f)}}{(1-p^2_L)\sigma_f}\bigg]
=\frac{k_B T}{eE
\sigma_s} \ln \frac{1+z_1}{1-z_1}}\\
{\displaystyle
\frac{p_R({\tilde G}_{\uparrow}+{\tilde G}_{\downarrow})}{2{\tilde
G}_{\uparrow}{\tilde G}_{\downarrow}}
-\frac{{\tilde G}_{\uparrow}-{\tilde G}_{\downarrow}}{2{\tilde
G}_{\uparrow}{\tilde G}_{\downarrow}}
+\Big[\alpha(x_0)-p_R\Big]\bigg[\frac{{\tilde G}_{\uparrow}+{\tilde
G}_{\downarrow}}
{2{\tilde G}_{\uparrow}{\tilde G}_{\downarrow}}+
\frac{2L^{(f)}}{(1-p^2_R)\sigma_f}\bigg]
=\frac{k_B T}{eE
\sigma_s} \ln \frac{1-z_2}{1+z_2}}
\end{array}\right.,
\end{equation}
where $p_L$ and $p_R$ are the spin polarization in the left and right
magnets, respectively, and
\begin{subequations}
\begin{eqnarray}
z_1&=&\frac{eE}{k_B T} \frac{\alpha(x_0)(L_u+L_d)-\alpha(0)(L_u e^{x_0/L_u}
+L_d e^{-x_0/L_d})}{e^{x_0/L_u}-e^{-x_0/L_d}},\\
z_2&=&\frac{eE}{k_B T} \frac{\alpha(x_0)(L_u e^{-x_0/L_u}+L_d e^{x_0/L_d})
-\alpha(0)(L_u
+L_d)}{e^{x_0/L_d}-e^{-x_0/L_u}}.
\end{eqnarray}
\end{subequations}
\end{widetext}

When $\alpha(0)$ and $\alpha(x_0)$ are known, we can express
the spin polarization of current in the semiconductor ($0<x<x_0$) as
\begin{eqnarray}
\alpha(x)&=&\alpha(0)\frac{e^{-(x-x_0)/L_d}-e^{(x-x_0)/L_u}}{e^{x_0/L_d}-e^{-x_0/L_u}}\nonumber\\
&-&\alpha(x_0)\frac{e^{-x/L_d}-e^{x/L_u}}{e^{x_0/L_u}-e^{-x_0/L_d}},
\end{eqnarray}
and the spin polarization of density in the semiconductor ($0<x<x_0$) as
\begin{eqnarray}
P(x)&=&-\alpha(0)\frac{eE}{k_B T}\frac{L_ue^{-(x-x_0)/L_d}+L_d
e^{(x-x_0)/L_u}}{e^{x_0/L_d}-e^{-x_0/L_u}}\nonumber\\
&+&\alpha(x_0) \frac{eE}{k_B T}\frac{L_u e^{-x/L_d}+L_d
e^{x/L_u}}{e^{x_0/L_u}-e^{-x_0/L_d}}.
\end{eqnarray}

Figure 5 depicts the spin injection efficiency at the left interface
($\alpha_0$)
as a function of electric field for the parallel
configuration $p_L=p_R=p_f$ and the anti-parallel configuration
$p_L=-p_R=p_f$.
For structures with transparent interfaces ($G^{-1}_{\uparrow}=
G^{-1}_{\downarrow}=\tilde{G}^{-1}_{\uparrow}=\tilde{G}^{-1}_{\downarrow}=0$),
we see that the spin injection in both configurations
can be enhanced by orders of magnitude
by increasing the field.
In the low-field regime, spin injection can only be achieved in structures
with the parallel
configuration.  In the high-field regime, spin injection from the left
magnet into the
semiconductor for both configurations are the
same, indicating that only the magnet from which carriers are injected  is
important to the
spin injection efficiency.

For structures with  transparent interfaces, spin polarization in the
semiconductor
$\alpha(x)$ can be expressed in compact forms in both the low- and
high-field
regimes. In fact,
in the low-field regime, where $x_0 \ll L_u, L_d$,  we
reproduce Eq. (7) in Ref. \onlinecite{schmidt},
\begin{equation}
\alpha(x)=\alpha=\bigg[\frac{2L^{(f)}}{(1-p^2_f)\sigma_f}+\frac{x_0}{\sigma_s}\bigg]^{-1}
\frac{(p_L+p_R)L^{(f)}}
{(1-p^2_f)\sigma_f},
\label{zero_e3}
\end{equation}
for $0<x<x_0$.
The above expression indicates that the spin injection strongly depends on
the relative orientation
of the two ferromagnetic metals:
For the parallel configuration,  $\alpha(x)$ is finite; whereas for the
anti-parallel
configuration, spin injection is not possible.
In this limit, the left magnet  and the right one are equally important in
determining spin injection.
The magnitude of spin injection $\alpha$ is determined by the ratio of
the effective resistances in the magnet ($L^{(f)}/\sigma_f$) and
in the semiconductor ($x_0/\sigma_s$). For typical device parameters
$L^{(f)}/\sigma_f \ll x_0/\sigma_s$, and the spin injection efficiency
would be  extremely small due to this resistance mismatch.
The spin polarization of density in this limit can be written as
$P(x)=-\alpha \Big(\frac{e E}{k_B T}\Big)^2 L_u L_d$.
At the zero-field limit, the density difference between up-spin and
down-spin
electrons vanishes.

On the other hand, in the high-field regime, $L_u$ and $L_d$ can differ by
orders of magnitude, and usually
$L_u \ll x_0 \ll L_d$. We find that in this limit, for $0 < x <x_0$,
\begin{equation}
\alpha(x) =\alpha=\Bigl[\frac{L^{(f)}}{(1-p^2_f)\sigma_f}
+\frac{L_u}{\sigma_s}\Bigr]^{-1} \frac{p_L
L^{(f)}}{(1-p^2_L)\sigma_f},
\label{ex_fnf}
\end{equation}
and
$P(x)=\alpha |e E|L_u/k_B T$, which is  close to $\alpha$ in the
strong-field limit.
We see that in the high-field regime,
spin injection is locally determined by the magnet from which carriers
are injected, and the ``remote'' magnet that collects the current becomes
irrelevant
in determining the spin injection efficiency.
Moreover, the effective resistance of the semiconductor to be compared
with that of the magnet ($L^{(f)}/\sigma_f$) is $L_u/\sigma_s$
rather than $x_0/\sigma_s$. In this regime, the spin polarization of current
and the spin polarization
of density have similar amplitude.
Thus in the high-field regime, according to Eq. (\ref{ex_fnf}),
the spin injection behavior in the sandwiched FM/NS/FM structure
would be  the same as  in a simpler FM/NS structure.
The reason that the second magnet becomes unimportant for spin injection in
the high-field
regime is that the influence of the second magnet on spin transport in the
semiconductor
is localized within the up-stream length ($L_u$) from the magnet.
$L_u$ decreases with increasing field and would be much shorter than $x_0$
in the high-field regime.

In Fig. 5, we also plot spin injection as a function of the electric field
in structure with two identical spin-selective interfacial barriers between
the
magnet and the semiconductor. In the calculations it is
assumed that
$G_{\uparrow(\downarrow)}=\tilde{G}_{\uparrow(\downarrow)}\ne 0$ for the
parallel configuration, and
$G_{\uparrow(\downarrow)}=\tilde{G}_{\downarrow(\uparrow)}\ne 0$ for the
anti-parallel configuration.
We see that the electric field substantially enhances spin injection
in structures with spin-selective interfacial barrier. In the high-field
regime, as in the transparent case, spin injection from the left magnet into
the semiconductor
does not depend on the orientation of the right magnet.

\begin{figure}
\vspace{10pt}\includegraphics[width=7cm]{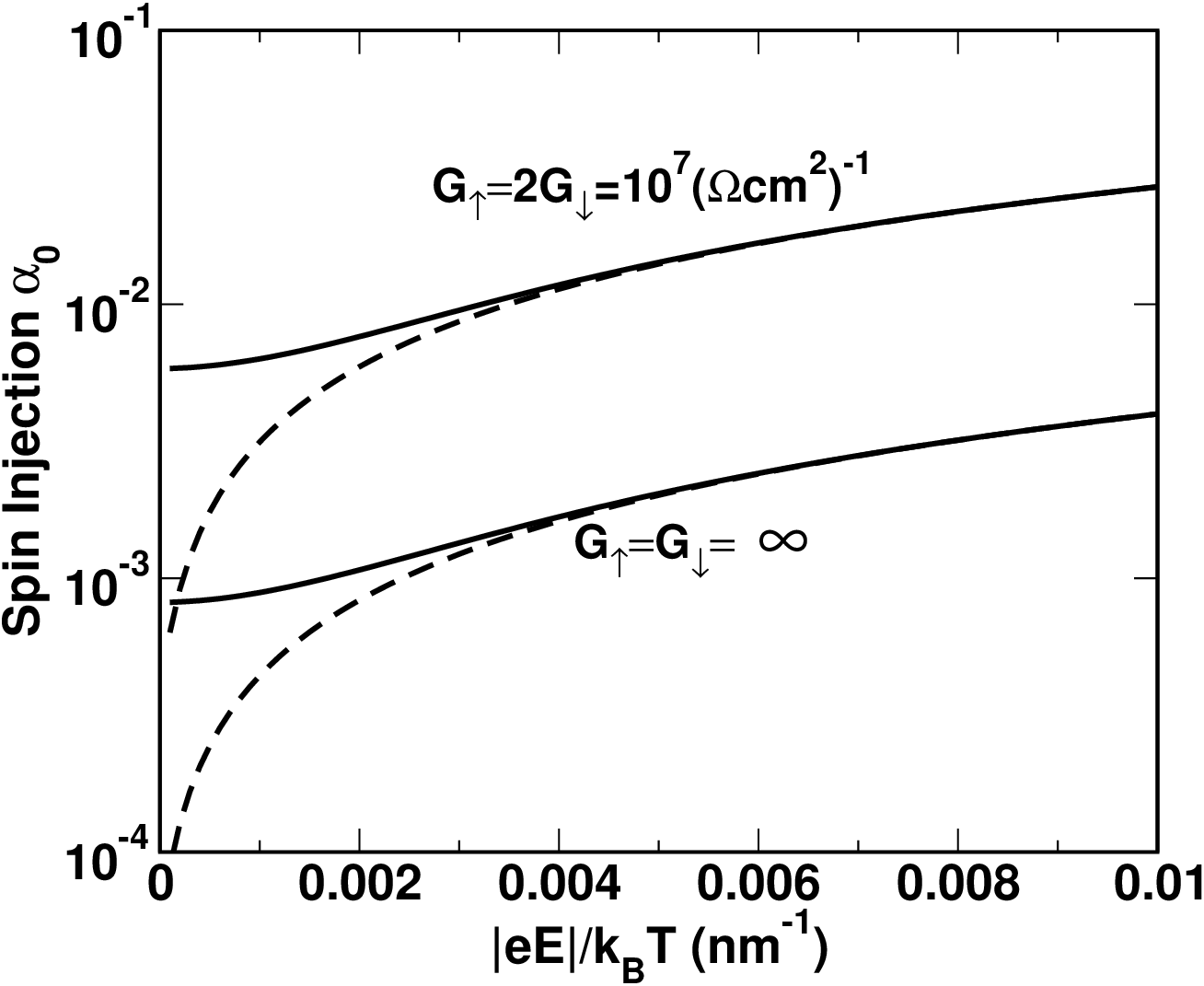}
\caption{Spin injection efficiency $\alpha_0$ as a function of electric
field. Solid and dashed lines correspond to parallel configuration ($p_L
=p_R=p_f$) and anti-parallel configuration ($p_L=-p_R=p_f$), respectively.
The lower curves are for structure with transparent interface. The
upper curves are for structure with a spin-selective
interfacial barrier, $G_{\uparrow}=2G_{\downarrow}=10^7~(\Omega$
cm$^2)^{-1}$.
Other parameters are $p_f=0.5$, $L^{(f)}=60$ nm, $x_0=1~\mu$m,
$L^{(s)}=2~\mu$m, and $\sigma_f=100~\sigma_s=10^3~(\Omega$ cm)$^{-1}$.}
\vspace{10pt}
\end{figure}

\section{Spin injection in FM/NS/NS structures}

In semiconductor spin injection structures,
a highly doped
nonmagnetic semiconductor (NS$^+$)
is often
placed near the magnet interface
to overcome the Schottky barrier between a magnet and
a semiconductor. This configuration  is also intrinsic
to FM/InAs, where  densely occupied surface states form at the interface.
Here we consider an injection structure that
comprises a semi-infinite magnet ($x <0$),
a finite nonmagnetic semiconductor with conductivity $\sigma_s$ and carrier
concentration
$n_0$ ($0<x<x_0$), and a semi-infinite nonmagnetic semiconductor
with conductivity $\tilde{\sigma}_s$ and carrier concentration
$\tilde{n}_0$
($x>x_0$).
The electrochemical potentials in the magnet satisfy Eq.
(\ref{e_potential}), and
the electron densities in semiconductors satisfy Eqs. (\ref{nondeg}) and
(\ref{neutral}). The general solution
of the electrochemical potentials in the magnet and the semiconductors
can be written as
\begin{equation}
\frac{1}{eJ}\left ( \begin{array}{c} \mu_{\uparrow}\\
\mu_{\downarrow}
\end{array} \right )
=\frac{x}{\sigma^f_{\uparrow}+\sigma^f_{\downarrow}} \left (
\begin{array}{c} 1\\ 1
\end{array} \right )+ C  \left
( \begin{array}{c}1/\sigma^f_{\uparrow}\\ -1/\sigma^f_{\downarrow}
\end{array} \right )e^{x/L^{(f)}}
\end{equation}
for $x<0$,
\begin{eqnarray}
\mu_{\uparrow(\downarrow)}&=&k_B T \ln \bigg[1+(-)\frac{2A_0e^{-x/L_d}+2A_1
e^{(x-x_0)/L_u)}}{n_0}\bigg]\nonumber\\
&+&\frac{e Jx}{\sigma_s} -B_0
\end{eqnarray}
for $0<x<x_0$, and
\begin{equation}
\mu_{\uparrow(\downarrow)}=k_B T \ln
\bigg[1+(-)\frac{2A_2e^{-(x-x_0)/\tilde{L}_d}}{\tilde{n}_0}\bigg]+\frac{e
Jx}{\tilde{\sigma}_s} -B_1
\end{equation}
for $x>x_0$.

Here we consider structures without interface resistance
between two semiconductor regions,
and the boundary conditions are given by Eq. (4.5a-f) with
$\tilde{G}^{-1}_{\uparrow}=\tilde{G}^{-1}_{\downarrow}=0$.
These equations completely determine the six unknowns, $A_i$ ($i=0,1,2$),
$B_i$ ($i=1,2$), and $C$, in expressions of Eqs. (5.1)-(5.3).

We obatin an equation for $z=\alpha(x_0)$, the spin polarization of
current at the interface between
the two semiconductors,
\begin{widetext}
\begin{eqnarray}
\frac{G^{-1}_{\uparrow}-G^{-1}_{\downarrow}}{2}+\frac{p_f
(G^{-1}_{\uparrow}+G^{-1}_{\downarrow})}{2}&-&
\frac{k_B T}{eJ}\ln \frac{-k_B T\tilde{\sigma}_s/eJ\tilde{L}_u+(a+b)z}{-k_B
T\tilde{\sigma}_s/eJ\tilde{L}_u-(a+b)z}\nonumber\\
&=&\bigg[\frac{2L^{(f)}}{(1-p^2_L)\sigma_f}+\frac{G^{-1}_{\uparrow}+G^{-1}_{\downarrow}}{2}\bigg]
\bigg[\Big(\frac{\tilde{L}_u\sigma_s}{L_u\tilde{\sigma}_s}a-\frac{\tilde{L}_u\sigma_s}{L_d\tilde{\sigma}_s}b\Big)z-p_f\bigg],
\end{eqnarray}
\end{widetext}
where $J=\sigma_s E = \tilde{\sigma}_s \tilde{E}$, and
\begin{eqnarray*}
a&=&\bigg[\sigma_s\Big(\frac{1}{L_d}+\frac{1}{L_u}\Big)\bigg]^{-1}
\Big(\frac{\tilde{\sigma}_s}{\tilde{L}_d}+\frac{\sigma_s}{L_u}\Big)e^{x_0/L_d},\\
b&=&\bigg[\sigma_s\Big(\frac{1}{L_d}+\frac{1}{L_u}\Big)\bigg]^{-1}
\Big(\frac{-\tilde{\sigma}_s}{\tilde{L}_d}+\frac{\sigma_s}{L_d}\Big)e^{-x_0/L_u}.
\end{eqnarray*}

Figure 6 illustrates the spin injection efficiency $\alpha(x_0)$
as a function of the total electric current $J$. We see that in the
low-field
regime, the conductivities of
both semiconductors are important to determine spin injection.
As the total effective resistance
($x_0/\sigma_s+\tilde{L}^{(s)}/\tilde{\sigma}_s$)
of semiconductors decreases, the spin injection efficiency increases.
With increase of the current or the field, spin injection can be enhanced
considerably.
Moreover, in the strong-field limit, the spin injection efficiency
will be determined by the total current flowing into the semiconductors.

In fact,  at the zero-field limit, where $L_u= L_d=L^{(s)}$, $\tilde{L}_u=
\tilde{L}_d=\tilde{L}^{(s)}$, and $x_0 \ll L^{(s)}$,
$\alpha(x_0)$ can be written as a simpler form,
\begin{widetext}
\begin{equation}
\alpha(x_0)=\bigg[
\bigg(\frac{L^{(f)}}{(1-p^2_f)\sigma_f}+\frac{G_{\uparrow}+G_{\downarrow}}
{4G_{\uparrow}G_{\downarrow}}\bigg)
\Big(1+\frac{\sigma_s}{\tilde{\sigma}_s}\frac{x_0\tilde{L}^{(s)}}{(L^{(s)})^2}\Big)
+\frac{x_0}{\sigma_s}+\frac{\tilde{L}^{(s)}}{\tilde{\sigma}_s}\bigg]^{-1}
\bigg[\frac{L^{(f)}p_f}{(1-p^2_f)\sigma_f}+\frac{G_{\uparrow}-G_{\downarrow}}
{4G_{\uparrow}G_{\downarrow}}\bigg],
\end{equation}
\end{widetext}
which shows that the ratio between the total effective resistance of the two
semiconductors,
$x_0/\sigma_s+\tilde{L}^{(s)}/\tilde{\sigma}_s$,
and that of the magnet, $L^{(f)}/\sigma_f$, together with
the spin-selective interfacial barrier between the ferromagnet and the semiconductor,
determines the spin injection
efficiency. On the other
hand,
at the strong-field limit, where  $L_u=k_B T/|eE|$, $\tilde{L}_u \ll
\tilde{L}_d$ and $L_u \ll x_0 \ll L_d$,
$\alpha(x_0)$ can be expressed in a even more compact form,
\begin{eqnarray}
\alpha(x_0)&=&\bigg[ \frac{L^{(f)}}{(1-p^2_f)\sigma_f}
+\frac{k_B T}{|eJ|}+\frac{G_{\uparrow}+G_{\downarrow}}
{4G_{\uparrow}G_{\downarrow}}\bigg]^{-1}\nonumber\\
&\times&\bigg[\frac{L^{(f)}p_f}{(1-p^2_f)\sigma_f}+\frac{G_{\uparrow}-G_{\downarrow}}
{4G_{\uparrow}G_{\downarrow}}\bigg],
\end{eqnarray}
which indicates that in FM/NS/NS structures spin injection
is controlled by the total current flowing into the semiconductors and a
distinction between the two semiconductors becomes unimportant.

\begin{figure}
\vspace{10pt}\includegraphics[width=7cm]{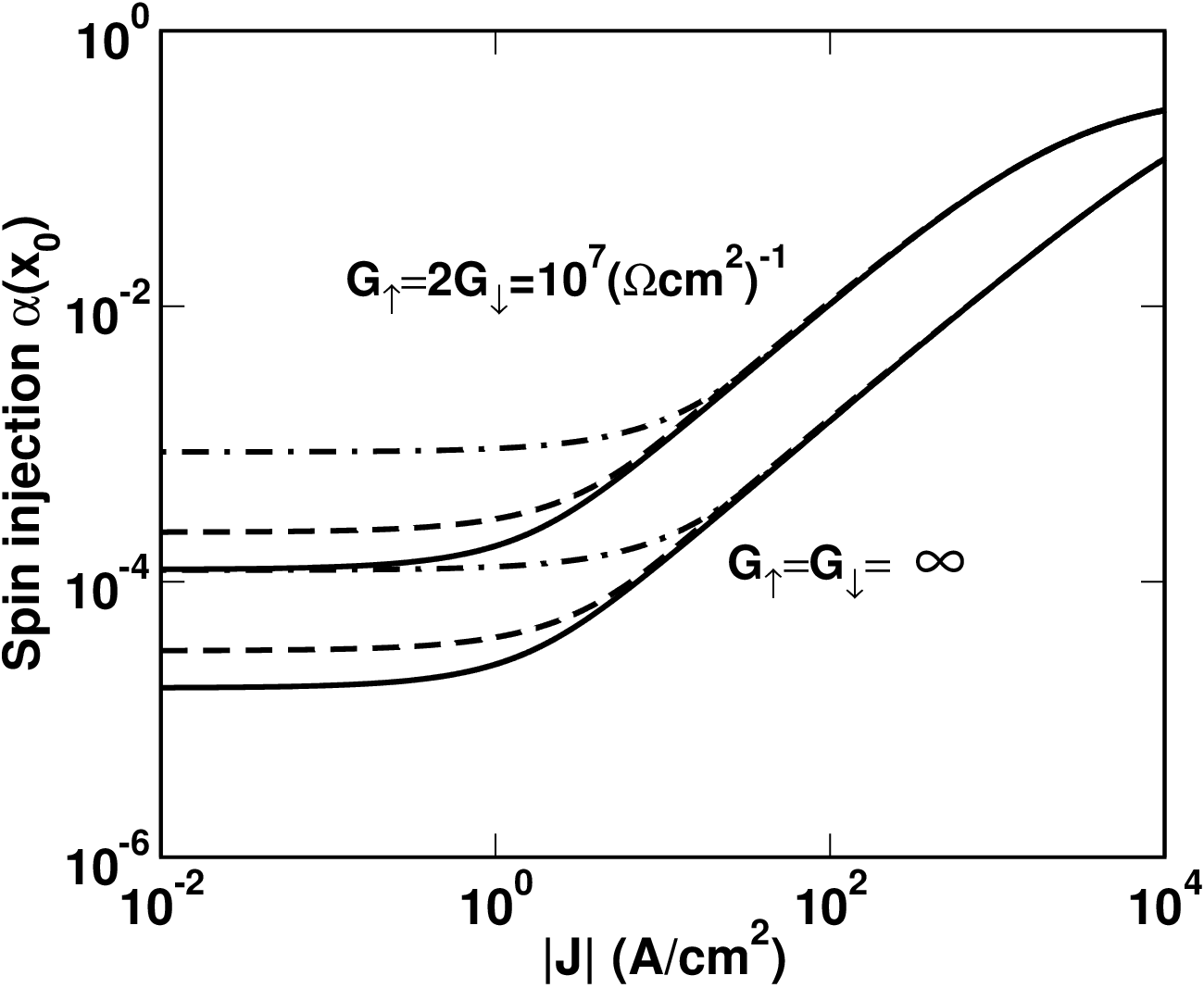}
\caption{Spin injection efficiency as a function of total electric
current. Solid, dashed, and dot-dashed lines correspond
to $(\sigma_s,\tilde{\sigma}_s)=(10,1)$, (1,10), and (10,10)
$(\Omega$ cm)$^{-1}$.
The lower curves are for structure with transparent interface. The
upper curves are for structure with a spin-selective
interfacial barrier, $G_{\uparrow}=2G_{\downarrow}=10^7~(\Omega$
cm$^2)^{-1}$.
Other parameters are $p_f=0.5$, $L^{(f)}=60$ nm,
$L^{(s)}=\tilde{L}^{(s)}=2~\mu$m, and $\sigma_f=10^3~ (\Omega$ cm)$^{-1}$.}
\vspace{10pt}
\end{figure}

\section{Field-dependent magnetoresistance in MS/NS/MS structures}

In MS/NS/MS structures a strong positive magnetoresistance effect
has been observed at low temperatures.\cite{magnetoresistance} As
the applied magnetic field is changed from 0 to $\sim 2~{\rm T}$
the spin polarization in a magnetic semiconductor can change at
low temperatures from 0 to $\sim 100$\%.

Electric-field dependence of the magnetoresistance
can be expected based on the different spin injection
behaviors in the high-field regime and in the low-field regime
discussed in Sec. IV. In the low-field regime, the densities of up-spin and
down-spin electrons in a nonmagnetic semiconductor
remain the same even in the presence of a fully
spin-polarized current. Thus in the semiconductor only {\em half} of
electrons contribute to the conductance if the current is 100\%
spin-polarized,
and the resistance of the semiconductor should be twice of that for a
unpolarized
current.
Hence the semiconductor resistance strongly depends on the spin polarization
of the current
in the low-field regime.
In the high-field regime, however, the spin polarization of density $P$
is close to the spin polarization of current $\alpha$. Therefore if
the current is 100\% spin-polarized, the electron density would be also
fully spin-polarized, and {\em all} electrons would contribute to the
conductance.
Hence in the high-field regime,
the semiconductor resistance is only weakly dependent on the spin
polarization,
and the magnetoresistance vanishes.

Here we calculate how the magnetoresistance depends on the
electric field. The magnetic semiconductor
we consider here is degenerate and its spin transport
is described by Eq. (\ref{e_potential}).  Thus all results obtained in Sec.
IV
are also applicable to the MS/NS/MS structures.
We also assume that the interfaces between the two materials are
transparent, i.e., zero interface resistance. The magnetizations of the
two magentic semiconductors are identical and parallel,
$p_L=p_R=p ({\bf H})$, which are
zero in the absence of external magnetic field and finite for a given
external magnetic field {\bf H}. The
resistance of the nonmagnetic semiconductor, $R$, would depend on the
spin polarization in the magnetic semiconductors because of spin
accumulation at the heterostructure interfaces, and is therefore
also a function of the external magnetic field.

The resistance of the nonmagnetic semiconductor can be calculated via
\begin{equation}
R \equiv \frac{\mu_0(x_0)-\mu_0(0)}{eJ},
\end{equation}
which certainly would be $R(0)=x_0/\sigma_s$
when attached to unpolarized
magnetic semiconductors (zero magnetic field).
The resistance in the presence of spin-polarized
magnetic semiconductors with $x_0 \ll L^{(s)}$ can be expressed as
\begin{equation}
R({\bf H}) \simeq x_0/\sigma_s+\frac{L^{(m)}p({\bf H})}{[1-p^2({\bf
H})]\sigma_m}\Big[2p({\bf H})-\alpha(0)-\alpha(x_0)\Big].
\end{equation}
Here $\alpha(0)$ and $\alpha(x_0)$ are the spin polarization of current at
the left and right interfaces,
$L^{(m)}$ the spin diffusion length in the magnetic semiconductor, and
$\sigma_m$
the conductivity of the magnetic semiconductor.
In the low-field regime, according to Eq. (\ref{zero_e3}), we find that the
magnetoresistance
\begin{equation}
\frac{\Delta R}{R}\equiv \frac{R({\bf H})-R(0)}{R(0)}=p^2({\bf H})
\Bigg(1+\frac{[1-p^2({\bf H})]\sigma_m x_0}{2\sigma_s L^{(m)}}\Bigg)^{-1}.
\end{equation}
This can be significant for an MS/NS/MS structure because of the
large $p({\bf H})$ [$p({\bf H}) \sim 1$] and small conductivity
($\sigma_m \sim \sigma_s$) in magnetic semiconductors.
We would like to point out, however, that
for ferromagnetic metal/semiconductor/ferromagnetic metal structures,
the magnetoresistance is usually too weak to detect due to the conductivity
mismatch
between metals and semiconductors ($\sigma_m \gg \sigma_s$).

In the high-field regime, by using Eq. (\ref{ex_fnf}) we express the
magnetoresistance as
\begin{equation}
\frac{\Delta R}{R}=\frac{2p^2({\bf H})}{x_0}
\Bigg(\frac{1}{L_u}+\frac{[1-p^2({\bf H})]\sigma_m
}{L^{(m)}\sigma_s}\Bigg)^{-1}.
\end{equation}
We see from the above expression that the effect of the electric field on
the magnetoresistance
can be described in terms of the field-induced up-stream spin diffusion
length. Increasing
the electric field will decrease the magnetoresistance  because $L_u$
decreases with the electric field.
Figure 7 illustrates the magnetoresistance as a function of electric
field for an MS/NM/MS structure. We see that
with increasing electric field  the magnetoresistance diminishes.
For example, if the parameters are chosen as follows:
$x_0=1~\mu$m, $L^{(m)}=60$ nm,
$L^{(s)}=2~\mu$m, $\sigma_m=\sigma_s$, and
$p({\bf H})=0.99$, the magnetoresistance $\Delta R/R$
decreases from 83\% at zero field to 5.3\% at $|eE|/k_B T=0.05$ nm$^{-1}$,
or $E=125$ V/cm for $\sigma_s=10~(\Omega$ cm)$^{-1}$ at $T=3$ K.
\begin{figure}
\vspace{10pt}\includegraphics[width=7cm]{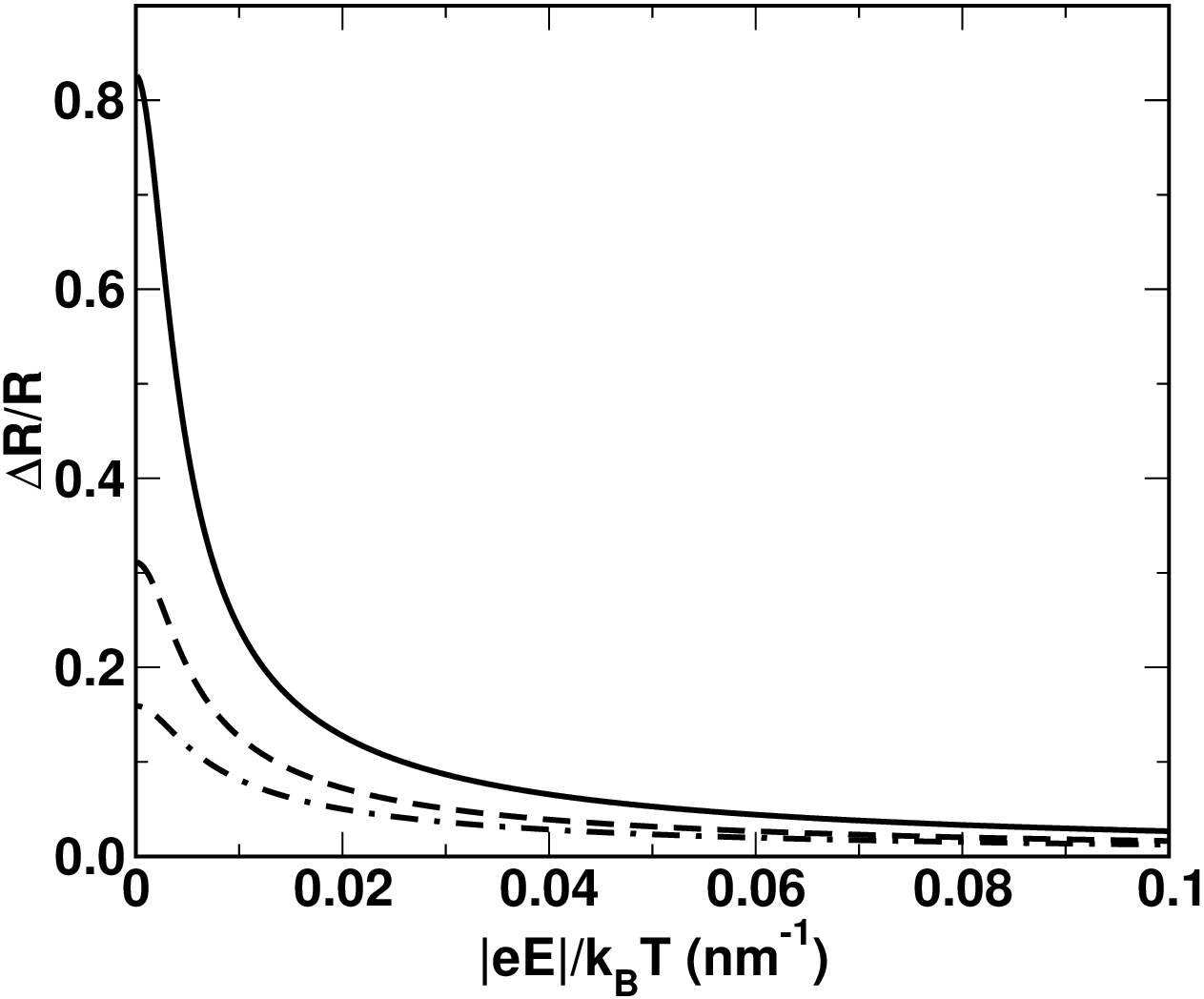}
\caption{Magnetoresistance $\Delta R/R$
as a function of electric
field. Solid, dashed, and dot-dashed lines correspond to
$p({\bf H})=0.99$, 0.9, and 0.8, respectively.
Other parameters are  $x_0=1~\mu$m, $L^{(m)}=60$ nm,
$L^{(s)}=2~\mu$m, and $\sigma_m=\sigma_s$.}
\vspace{10pt}
\end{figure}

\section{concluding remarks}

In summary,
we have derived a general drift-diffusion equation for spin polarization
in both degenerate and nondegenerate systems
by consistently taking into account
electric-field effects. We have demonstrated that as a system changes
from degenerate to nondegenerate, the electric field becomes
more and more important in spin transport.
We have identified a high-field diffusive regime in nondegenerate
semiconductors
which has no analogue
in metals. In this regime, there are two distinct
spin diffusion lengths, i.e.,  up-stream  and down-stream
spin diffusion lengths.

We have applied this more general drift-diffusion equation for spin
polarization
to several typical injection structures encountered in
semiconductor spintronic devices to study the spin injection behavior
in these structures and the effects of electric fields.
The high-field description of the spin transport in semiconductors
predicts that the electric field can effectively enhance
spin injection
from a ferromagnet  into a semiconductor.
For structures with a spin-selective interfacial barrier we find that
the electric field further enhances spin injection substantially.
The combination of the field enhancement and the interface enhancement of
spin injections may help us to obtain a comprehensive understanding
of observed large spin injection  in a variety of FM/NS, FM/NS/FM, and
FM/NS/NS structures
and current-dependent spin injection.

The consequences of spin injection into a semiconductor in the
high-field regime are qualitatively different from those in the
low-field regime. A high-field injection creates a notable density
difference between up-spin and down-spin electrons and the spin
polarization of density has a similar value as the spin
polarization of current. In contrast a low-field injection  does
not create an appreciable density difference between up-spin and
down-spin electrons, and the spin polarization of current is quite
different from the spin polarization of density. One consequence
of these different spin injection behaviors is that for sandwiched
FM/NS/FM structures the high field destroys the symmetry between
the two magnets at low fields, where both magnets are equally
important to determine spin injection. The efficiency of spin
injection into semiconductors in the high-field regime is
``locally'' determined by the magnet from which carriers are
injected into the semiconductor and the magnet that collects the
carriers becomes irrelevant. Another consequence is that for
FM/NS/NS structures spin injection efficiency in the high-field
regime is only determined by the total injected electric current
and the distinction between the semiconductors becomes
unimportant.

We have also examined the electric-field effect on magnetoresistance in
MS/NS/MS structures. In the low-field regime, the magnetoresistance in an
MS/NS/MS structure
can be significant, as reported in Ref. \onlinecite{magnetoresistance}. With
increasing
electric field, the magnetoresistance diminishes quickly. The underlying
physics
is that in the high-field regime the spin polarization of density is similar
to
the spin polarization of current in the nonmagnetic semiconductor and {\em
all}
electrons contribute to the conductance in the presence of spin-polarized
current in contrast to {\em half} of electrons in the low-field regime
if the current is fully polarized.
Thus in the high-field regime, the resistance of the nonmagnetic 
semiconductor
only weakly depends on the spin polarization of current and the magnetic 
field,
giving rise to a diminishing
magnetoresistance effect in the high-field regime.

Our calculations in this paper present a broad picture of
electric field-dependent spin transport and spin injection
phenomena into semiconductors. Our theory
also provides physical insight into the field-induced enhancement of spin 
injection
as well as the electric field-dependent magnetoresistance
and suggests high-field injection as a simple way to
amplify spin injection into semiconductor spintronic devices.

\acknowledgments

This work was supported by DARPA/ARO DAAD19-01-0490.

\end{document}